%% file: main.tex
\documentclass{IEEEtran}
\usepackage{amsmath,amssymb,amsfonts}
\usepackage{graphicx}
\def\BibTeX{{\rm B\kern-.05em{\sc i\kern-.025em b}\kern-.08em
    T\kern-.1667em\lower.7ex\hbox{E}\kern-.125emX}}

\usepackage{mathptmx} 
\newcommand{\ignore}[1]{}
\usepackage{fancyhdr}
\usepackage[normalem]{ulem}
\usepackage[hyphens]{url}
\usepackage{microtype}
\usepackage[pass]{geometry}
\usepackage{multirow}
\usepackage{subcaption}
\usepackage{xspace}
\usepackage{color}
\usepackage{soul}

\usepackage[bookmarks=true,breaklinks=true,letterpaper=true,colorlinks,linkcolor=black,citecolor=blue,urlcolor=black]{hyperref}

\usepackage{booktabs}
\usepackage[sort,nocompress]{cite}

\newcommand{\NAME}{MASR\xspace}

\newcommand{\TABSEPsmall}{\hspace{0.5em}}

\fancypagestyle{firstpage}{
  \fancyhf{}
\setlength{\headheight}{50pt}

  \fancyhead[C]{\normalsize{To appear in 28$^{th}$ International Conference on Parallel Architecture and Compilation Techniques (PACT 2019)}}
}

\title{\NAME: A Modular Accelerator for Sparse RNNs \vspace{1em} \thanks{To appear in 28$^{th}$ International Conference on Parallel Architecture and Compilation Techniques (PACT 2019)}}
\author{Udit Gupta, Brandon Reagen, Lillian Pentecost, Marco Donato, Thierry Tambe, \\
Alexander M. Rush, Gu-Yeon Wei, David Brooks \\
ugupta@g.harvard.edu}

\begin{document}
\maketitle
\pagestyle{plain}


\begin{abstract}

  Recurrent neural networks (RNNs) are becoming the \textit{de facto}
  solution for speech recognition. RNNs exploit long-term
  temporal relationships in data by applying repeated, learned
  transformations.
  Unlike fully-connected (FC) layers
  with single vector matrix operations, RNN layers consist
  of \emph{hundreds} of such operations chained over time.
  This poses challenges unique to RNNs that are not found in convolutional neural networks (CNNs) or FC models, namely large dynamic activation.
  In this paper we present \NAME, a principled and modular architecture that accelerates bidirectional RNNs for on-chip ASR. \NAME is designed to exploit sparsity in both dynamic activations and static weights.
  The architecture is enhanced by a series of dynamic activation optimizations that enable compact storage, ensure no energy is wasted computing null operations, and maintain high \textit{MAC} utilization for highly parallel accelerator designs.
   In comparison to current state-of-the-art sparse neural network accelerators (e.g., EIE), \NAME provides 2$\times$ area 3$\times$ energy, and 1.6$\times$ performance benefits.
   The modular nature of \NAME enables designs that efficiently scale from resource-constrained low-power IoT applications to large-scale, highly parallel datacenter deployments.
\end{abstract}




\input{introduction}

\input{related}
\input{ds2_thierry}
\input{baseline}

\input{optimizations}

\input{arch}

\input{designspace}

\input{inefficiencies}

\input{discussion}
\input{conclusion}


\bibliographystyle{ieeetr}
\bibliography{main}

\end{document}

%% file: introduction.tex
\section{Introduction}

\begin{figure}[t!]
    \centering
    \includegraphics[width=\linewidth]{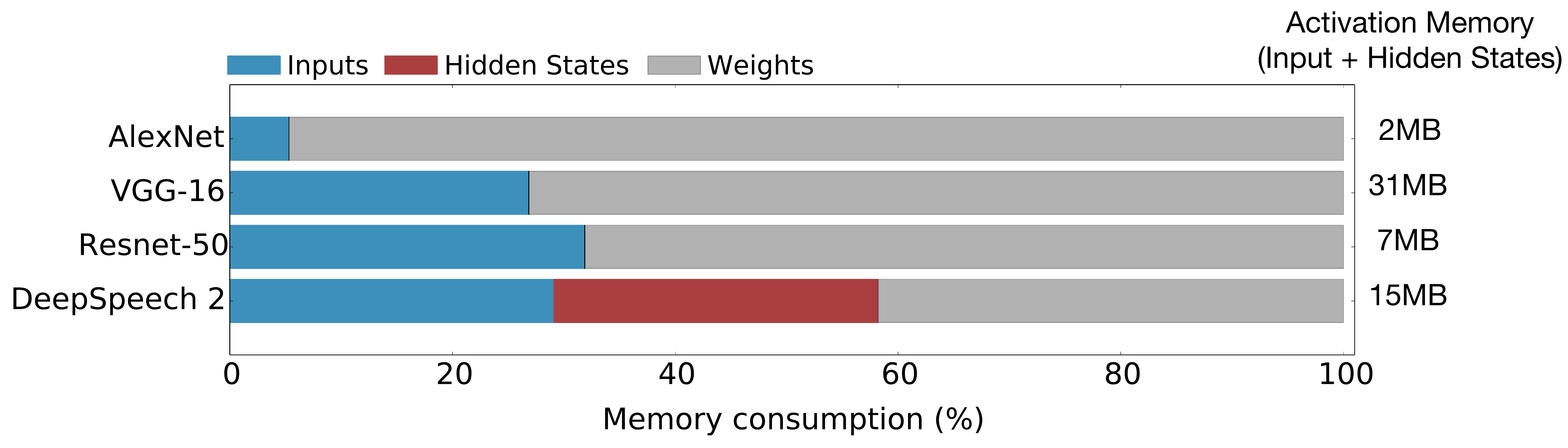}
    \vspace{-1.5em}
\caption{The memory footprint of activations is higher in Deep Speech~2 (DS2), an ASR RNN, than in standard CNNs.
Thus, to reduce storage costs of ASR RNNs, memory system optimizations are needed for \textit{both} activations and weights.}
\vspace{-1em}
\label{fig:models}
\end{figure}

Automatic speech recognition (ASR) is at the foundation of many popular services,
streamlining the human-machine interface~\cite{alexa, duplex}.
Recent advances in ASR have come from replacing
traditional methods based on Gaussian Mixture Models and Hidden Markov Models 
with deep learning, namely recurrent neural networks (RNNs).
RNNs learn relationships in time series data by
establishing a temporal context called the \textit{hidden state}---partial
predictions between time-adjacent neurons
that improve the interpretation of sequential data (e.g., spoken utterances).
Today, RNNs are the state-of-the-art solution for highly-accurate
ASR~\cite{ds2,sak2014long, Sak2015FastAA, googlevoicesearch}.
The hidden state of RNNs introduces a unique memory consumption problem that is addressed in this paper.
Figure~\ref{fig:models} compares the fraction of memory used by
activations and weights across four deep learning models.
Well-known, CNN-based image classification models
devote most of their memory resources to storing weights.
In contrast, nearly 60\% of the memory needed for Deep Speech~2 (DS2) ---a state-of-the-art,
RNN-based ASR model---is for activations (both inputs and hidden states), which consumes significant on-chip storage.
This does not preclude the issue of weights also consuming significant memory (14MB for Deep Speech 2).
These memory requirements are a result of ASR RNNs often using bidirectional layers --- inputs to each layer are processed twice (once forwards in time and once backwards) and
work over hundreds to thousands of time steps (i.e., 1 to 30 seconds)~\cite{libri}.
The hidden state size scales with the number of time steps, and separate weights are maintained for forward and backward passes.


Aggressive optimizations are needed to
reduce the memory costs of storing \textit{both} activations and weights, as well as the heavy processing load.
One promising solution is leveraging sparsity for storage and computational efficiency.
However, while many techniques for weight pruning and compression have been proposed, relatively little has been done to compress activations.
To improve computational efficiency, inferences can be computed directly on the sparse encoding.
Sparse processing allows the hardware to elide all null operations at the expense of introducing irregularity.
Irregularity leads to hardware inefficiency from low utilization,
and the optimal sparse encoding is application dependent.
In addition to not considering activation sparsity,
existing, CNN-centric solutions\cite{eie,scnn,cambricon,eyeriss}
are either not applicable to RNNs or perform poorly.
Enabling ubiquitous ASR requires accelerating RNNs with
algorithm-architecture co-design for sparse storage and efficient execution.

This paper presents \NAME:
A \underline{m}odular \underline{a}ccelerator to efficiently process \underline{s}parse \underline{R}NNs.
Through algorithm-architecture co-design,
\NAME achieves high hardware utilization while
never wasting area nor energy on superfluous computation.
To demonstrate the efficacy of the proposed technique,
we start by aggressively optimizing our baseline RNN
with knowledge distillation, language modeling, weight pruning, and quantization.
The key research contributions of \NAME fall into three categories:
a hardware accelerator that exploits sparsity in both weights and activations to skip null values in execution and storage,
a co-designed sparse encoding technique for both activations and weights that enables highly parallel architectures,
and a mechanism for dynamic load balancing to maximize hardware utilization.



\textbf{Sparse ASR RNN accelerator}
\textit{Algorithmic optimizations (i.e., knowledge distillation) and \NAME's co-designed micro-architecture exploit sparsity in weights and activations leading to improved performance, area, and energy by 14$\times$, 2$\times$, and 15$\times$ compared to a dense ASR RNN baseline.}

In order to reduce storage and computational burdens of ASR RNNs, activations must be sparse.
However, unlike CNNs, RNN activations (inputs and hidden states) are not typically sparse.
To achieve hidden state sparsity we use knowledge distillation~\cite{nips2014distill} to train RNNs with ReLU, at no loss in accuracy compared to GRU baselines with tanh.
Furthermore, we maintain input sparsity across layers by refactoring the batch normalization operation.
These modifications expose sufficient sparsity in RNNs to co-design our sparse activation-weight encoding.

\input{tables/relatedwork.tex}

\textbf{Sparse encoding}
\textit{\NAME's low-cost and scalable sparse encoding technique, provides a 2$\times$ area, 3$\times$ energy, and 1.6$\times$ performance benefit
relative to a start-of-the-art sparse DNN accelerator~\cite{eie}.
}

\NAME's sparse encoding format is co-designed with the underlying architecture to address both compute and storage bottlenecks.
Existing sparse encodings,
in addition to not compressing activations,
exhibit high meta-data costs stemming from encoding overheads.
\NAME proposes a binary-mask sparse encoding scheme for both weights and activations.
In storing bits rather than pointers, \NAME replaces expensive memory addressing with cheap bit-wise operations.

\textbf{Dynamic load balancing}
\textit{\NAME dynamically balances load from the irregular distribution of non-zero activations to improve performance by up to 30\% and achieve high MAC-utilization across a wide range of parallel design points.
}

Irregularity introduced by sparsity can lead to poor hardware utilization~\cite{isca2017tpu, NVDLA,cambricon}.
\NAME is designed to maximize utilization by,
(1) considering both intra- and inter-neuron parallelism, and
(2) employing a decoupled pipeline to separate the irregularity from sparsity from the computation of partials.
Once work is issued to the backend, the pipeline does not stall, regardless of the sparsity pattern.
The remaining source of low utilization arises from load imbalance ---  pipelines with more sparsity complete before others.
To improve hardware utilization we propose a dynamic load balancing
technique to re-distribute activations at run-time with negligible area and energy overheads.





\begin{figure}[!t]
\centering
\includegraphics[width=0.95\linewidth]{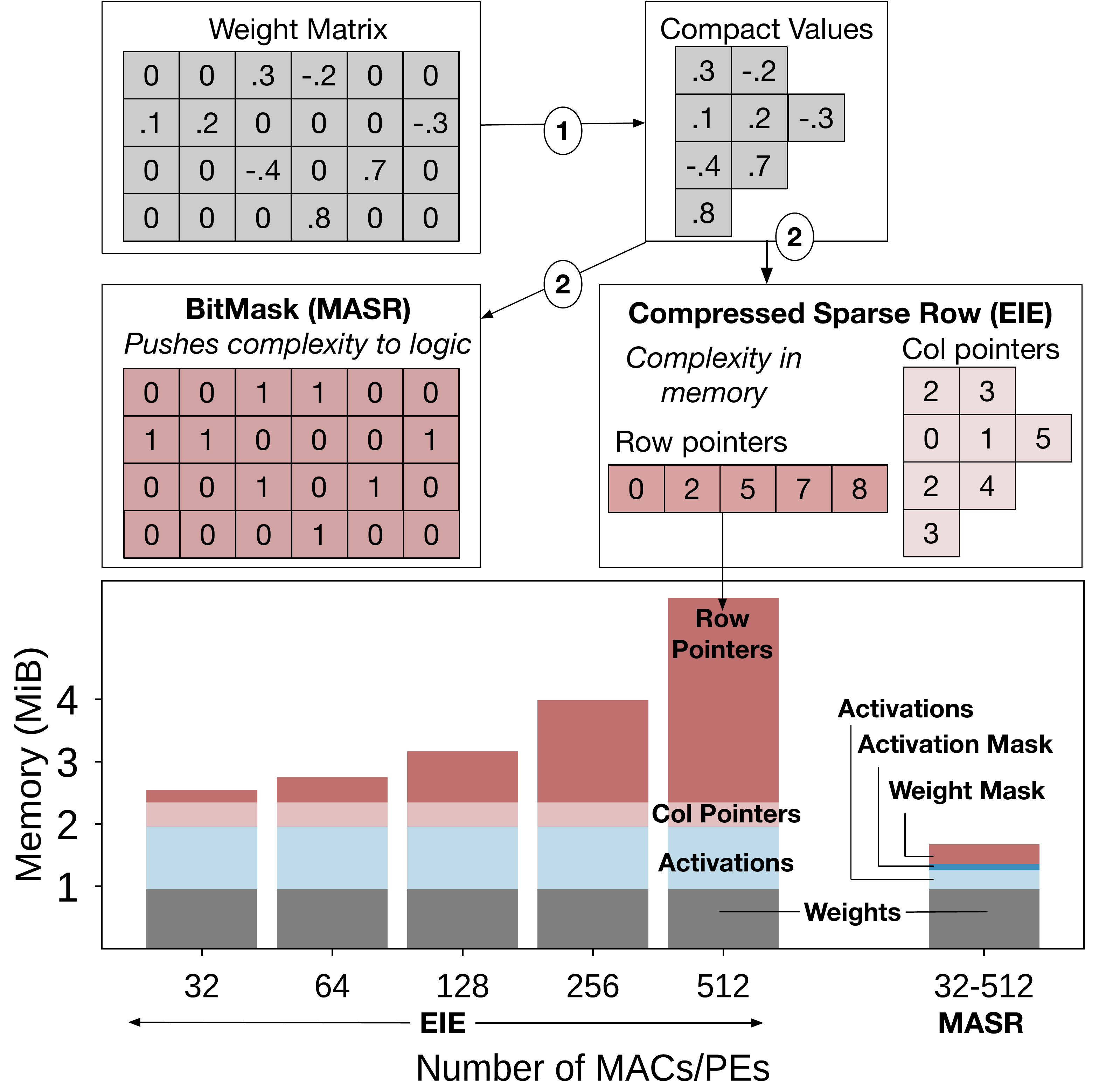}
\caption{Compared to compressed sparse row encoding (e.g., EIE~\cite{eie}), MASR's bit-mask encoding pushes the complexity in sparse encoding away from storing pointers to logic.
While storage for costly row pointers in EIE scales with the number of PEs, MASR's sparse encoding storage overhead is constant --- providing scalability.}
\label{fig:sparse_enc_motiv}
\vspace{-1.0em}
\end{figure}

%% file: tables/relatedwork.tex
\begin{table}[t!]
\small
\caption{Comparing \NAME to related work in terms of support for sparse execution and storage running RNNs. } 
\vspace{-1.0em}
  \begin{center}
    \begin{tabular}{c@{\TABSEPsmall}| c@{\TABSEPsmall}| c@{\TABSEPsmall}| c@{\TABSEPsmall}| c@{\TABSEPsmall}| c@{\TABSEPsmall}}
     \toprule
     & \multicolumn{2}{|c|}{\textbf{Sparse Weight}}
     & \multicolumn{2}{|c|}{\textbf{Sparse Act.}}
     & Dyn. Load \\ 

     & Exec. & Storage & Exec. & Storage & Balancing \\ \hline
     
     E-PUR~\cite{epur} & & & & & \\ \hline
     Minerva~\cite{zhu2018sparsenn} & & & x & & \\ \hline
     SparseNN~\cite{zhu2018sparsenn} & & & x & & \\ \hline
     Camb-X~\cite{cambricon} & x & x & & & \\ \hline
     ESE/EIE~\cite{eie} & x & x & x & & \\ \hline \hline
     MASR & x & x & x & x & x \\ \hline

   \end{tabular}
\end{center}
    \label{tab:related}
\vspace{-2em}
\end{table}

%% file: related.tex
\section{Related Work}

\textbf{Accelerating ASR RNNs}
Deep neural networks entail a general class of machine learning models that have been deployed across a wide set of applications and platforms~\cite{hpca2018amlfb,fathom,isca2017tpu,park2018deep}.  
Given their ability to achieve state-of-the-art accuracy in a broad range of applications, DNNs have gained a lot of attention from the architecture community.
However, much of the effort has been devoted to optimizing DNNs with only FC and CNN layers~\cite{diannao,dadiannao,eie,eyeriss,minerva,cambricon,scnn,cnvlutin,epur,circnn,isaac,DBLP:journals/corr/RhuOCPK17,venkataramani2017scaledeep,song2018prediction, chi2016prime, judd2016stripes, sharma2018bit, Ren:2019:AAC:3297858.3304076,kung2019packing,jin2019split}.
RNNs, used widely in ASR and natural language processing, pose unique challenges.
For instance, activations (inputs and hidden-states) generated at run-time comprise a higher fraction of the memory consumption in RNNs than in FC/CNNs (Figure~\ref{fig:models}).

Beyond accelerators for DNNs and CNNs, other work has investigated RNNs, and search algorithms for ASR and machine translation~\cite{li2019rnn, papamichael2018configurable, sharma2018bit, zhang2018towards, kwon2018maeri, Sivathanu:2019:AEP:3297858.3304072,gao2019tangram}. 
Shown in Table~\ref{tab:related}, E-PUR~\cite{epur} provides a hardware accelerator that maximizes weight locality in dense RNNs.
Similarly, the authors in \cite{congrnn,eugeniornn,drnn} leverage the temporal locality of dense RNNs to accelerate them on FPGAs. 
In contrast, MASR exploits sparsity in both weights and activations to further improve performance, area, and energy efficiency.

To accelerate ASR, the authors of~\cite{unfold} design  a memory-efficient Viterbi search accelerator.
This targets language models that are run after processing all timesteps and layers in the RNN.
However, with even large language models, state-of-the-art ASR models~\cite{ds1,ds2} spend over 90\% of their execution time on the RNNs (Section~\ref{sec:modelopt}), making RNNs the performance bottleneck and the focus of this paper.



\textbf{Exploiting sparsity for hardware efficiency}
Table~\ref{tab:related} compares \NAME to previous hardware accelerators based on their support for sparsity in weights and activations, and dynamic load balancing.
Typically, previous work either exploits sparsity in weights or activations, but not both~\cite{sze2017efficient, epur, minerva, fpga2017esehan, eyeriss, zhu2018sparsenn, cambricon} leaving key performance, area, and energy savings on the table.
DNN accelerators that do exploit sparsity in both weights and activations use dataflows and sparse encodings specific to CNNs~\cite{cnvlutin, scnn}. 
Thus, to highlight the key contributions made in this paper, we provide in depth comparisons to EIE~\cite{eie}. 

While EIE~\cite{eie} exploits sparsity in both weights and activations, it does not store activations in a compressed format.
Furthermore, EIE uses compressed-sparse row (CSR) encoding.
As shown in Figure~\ref{fig:sparse_enc_motiv}, CSR maintains separate row and column pointers to track non-zero weights.
Row-pointer storage scales with the number of hardware PEs, levying high memory costs in more parallel architectures.  
In contrast, \NAME uses a simpler sparse encoding that pushes the complexity of computing addresses for sparse parameters away from memory and into low-cost logic. 
This facilitates scaling the architecture to highly parallel designs (see Section~\ref{sec:sparsenc} for details).


\textbf{Load balancing for sparse neural networks}
Exploiting sparsity in weights and activations comes at the expense of introducing irregularity into an otherwise regular workload.
Irregularity leads to low hardware utilization from load imbalance.
Prior work considers pruning to statically balance weight sparsity~\cite{fpga2017esehan}.
However, we find that the main source of imbalance in RNNs is the distribution of non-zero activations.
Thus, \NAME exploits a novel dynamic load balancing technique that balances non-zero activations at run-time (Section~\ref{sec:inefficiency}).

%% file: ds2_thierry.tex
\begin{figure}[!t]
\centering
\includegraphics[width=\linewidth]{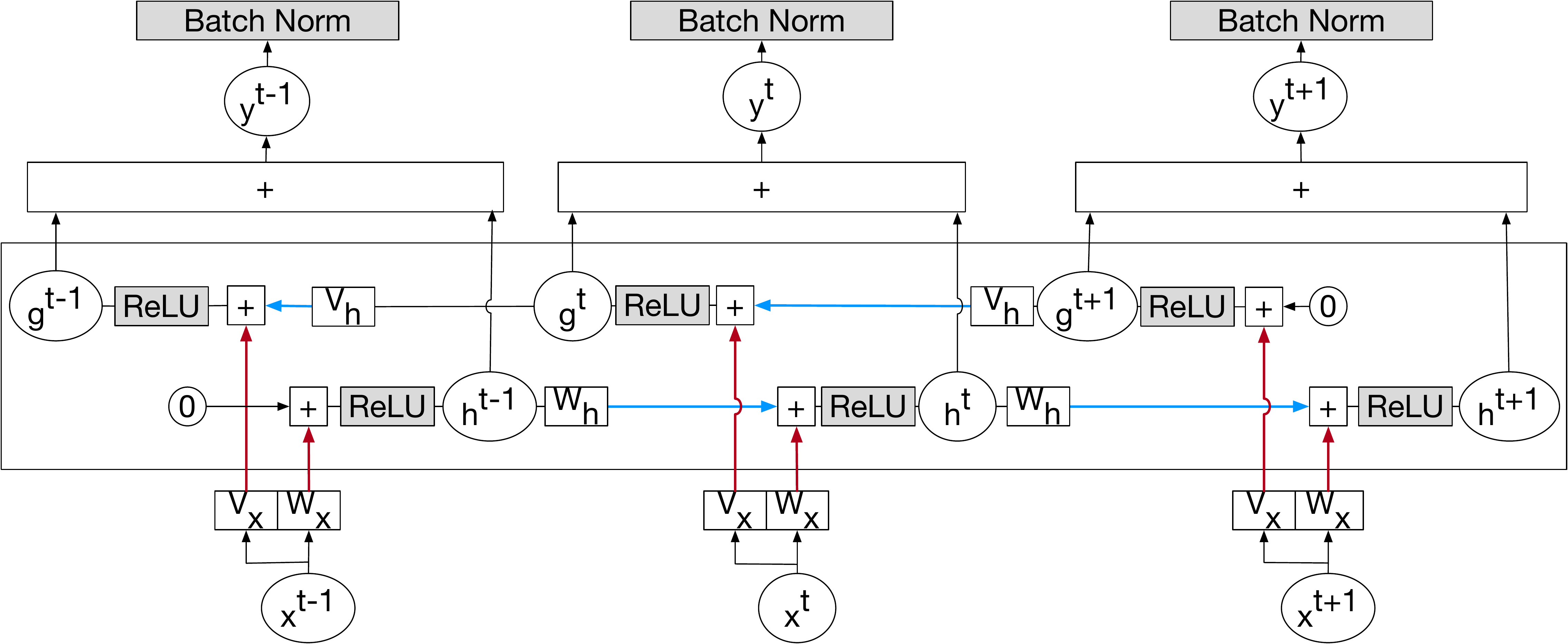}
\vspace{-0.5em}
\caption{Bidirectional RNN layer. $x^t$, $h^t$, and $g^t$ are
  the input and
  hidden states at time step $t$. 
  $W_x$, $W_h$, $V_x$, and $V_h$ are the forward and backward weights.}
\label{fig:rnn}
\vspace{-1.0em}
\end{figure}

\section{Automatic Speech Recognition} \label{sec:asr}

Automatic speech recognition (ASR) transcribes an input $x$ into text.
The input speech is represented as a discrete time series of continuous feature vectors $x^1, \ldots, x^T$  
derived from a spectrogram of power normalized audio clips.
Current state-of-the-art models for ASR rely heavily on deep learning for  acoustic modeling \cite{mohamed2012acoustic,hinton2012deep}.
Recently, RNNs have become the standard end-to-end deep learning approach for ASR ~\cite{ds2,sak2014long}. 
This section first provides an overview of RNNs and how they are used in ASR. 
We then simplify the neural networks to establish an efficient baseline RNN for ASR. 

\subsection{Recurrent Neural Networks}
Recurrent layers build 
context in time series data using hidden states to learn feature patterns in long-term data. 
There are three popular recurrent layers: vanilla RNN (hereafter referred to as RNN),
GRU \cite{chung2014empirical}, and LSTM \cite{hochreiter1997long}.
They differ in how new inputs are used to update hidden state.
RNNs use two sets of weights: one for inputs and one for hidden states.
GRUs and LSTMs expand upon RNNs with additional gated skip connections. 
These can improve accuracy by increasing expressiveness; however, the additional connections increase model size, where GRUs and LSTMs have 3$\times$ and 4$\times$ more weights than RNNs, respectively.


Recurrent layers can either be unidirectional or
bidirectional. Unidirectional layers update
hidden states based entirely on information from previous time steps.
Bidirectional layers maintain two separate hidden
states, one based on inputs from the past and one based on inputs from the future.  
While bidirectional layers can achieve higher accuracy, they require twice 
the number of parameters and operations. 

Figure~\ref{fig:rnn} illustrates a bidirectional RNN layer with ReLU 
activation and batch normalization. 
From the bottom, first a time-series input $x^t$ is transformed by matrices $W_x$ and $V_x$ to produce \textit{input intermediates} (red).
Hidden states $h^{t-1}$ and $g^{t+1}$ are transformed by matrices $W_h$ and $V_h$, respectively, to produce forward and backward \textit{hidden intermediates} (blue). 
New hidden states $h^t$ and $g^t$ are then computed by  passing the sum of the input and hidden intermediates through ReLU.
The sum of these hidden states is output as $y^t$. 




\subsection{Target Model: Deep Speech 2}
Deep Speech~2 (DS2)~\cite{ds2} is an industry and academic standard speech-to-text benchmark~\cite{mlperf}.
It directly maps input speech spectrograms to characters.
Table~\ref{tab:ds2} describes the architecture using an implementation based on GRUs.
First, a pair of CNN layers extract relevant features from the input spectrogram and reduce the length.
Next, bidirectional recurrent layers, which can either be GRU or RNN layers~\cite{ds2}, learn time-series context. 
Finally, a FC layer makes output predictions, 
a probability distribution over characters at each time step. 
This distribution can be combined with a language model to produce better transcribed text.

\input{tables/ds2_table.tex}





Our models for ASR are trained using the open-source DS2 implementation in PyTorch \cite{pytorch, seannaren,ctc} on the open-source LibriSpeech corpus \cite{libri}.
The GRU network (described in Table~\ref{tab:ds2})
has a word-error rate (WER) of 21.9, comparable to DS2 networks with a greedy decoder~\cite{ds2}
and the target for standardized speech-to-text benchmarks~\cite{mlperf}. 
%
The GRU layers make up over 99\% of the model's parameters. 
Thus, this paper focuses on optimizing the recurrent layers of DS2.





%% file: tables/ds2_table.tex

\begin{table}[t!]
\small
\caption{DS2 \cite{mlperf} model before optimizations (21.9 WER)} \vspace{-1.5em}
  \begin{center}
    \begin{tabular}{c@{\TABSEPsmall}| c@{\TABSEPsmall}| c@{\TABSEPsmall}| c@{\TABSEPsmall}}
     \toprule
      & Convolution &  Bidirectional GRU & Fully-connected \\ \cmidrule{1-4}
     Layers     & 2 & 5 & 1 \\ \cmidrule{1-4}
     Parameters & 250K & 38M & 20K \\ \hline
   \end{tabular}
\end{center}
    \label{tab:ds2}
\vspace{-2em}
\end{table}


%% file: baseline.tex
\subsection{An Efficient RNN Baseline}  \label{sec:modelopt}
\NAME builds on a very efficient baseline design that includes several previously-proposed 
optimization techniques to improve performance, on-chip area, and energy costs of DNNs. 
These techniques---knowledge distillation, language modeling, weight pruning, and quantization---were adapted for ASR RNNs.

\begin{figure}[!t]
\centering
\includegraphics[width=\linewidth]{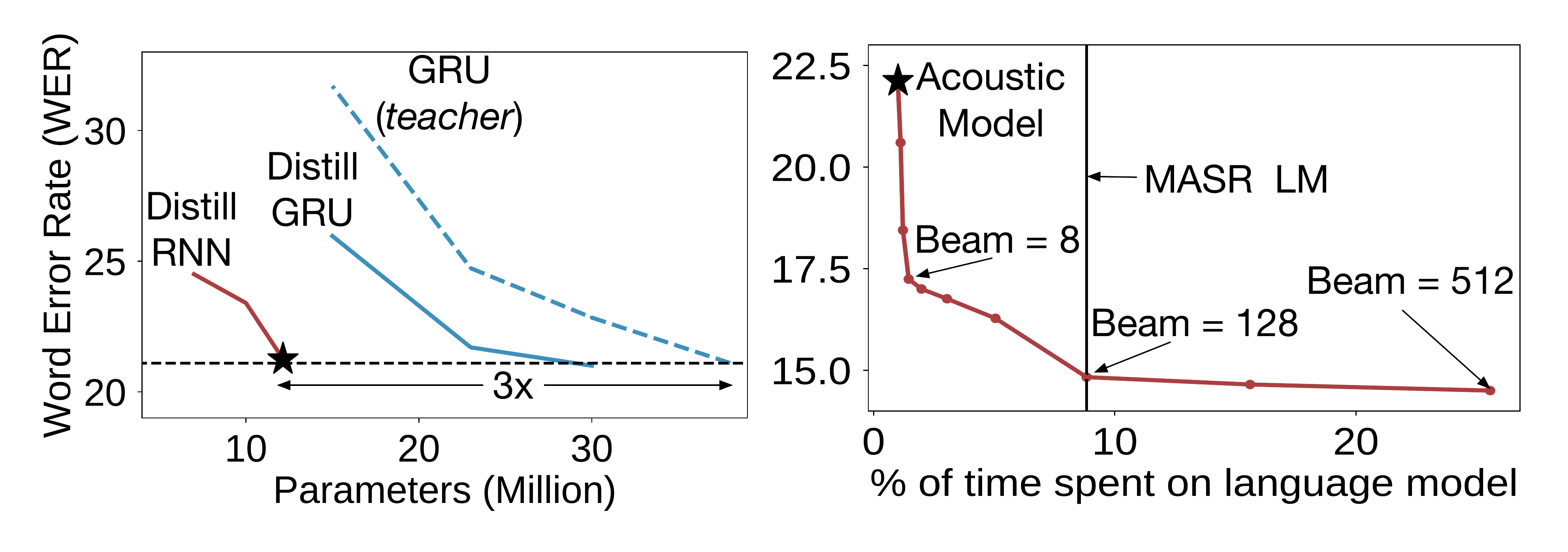}
\caption{ \textbf{Left}: After distillation, a 5-layer bidirectional RNN
  reaches the accuracy of a 5-layer   bidirectional GRU reducing the number of parameters by 3$\times$.
  \textbf{Right}: Language modeling reduces the WER from 22 down to 14.5 with a beam-width of 128. Language modeling accounts for only 10\% of CPU time.
  }
\label{fig:distill}
\vspace{-1.5em}
\end{figure}

\textbf{Knowledge Distillation} is a technique used to train a smaller, less complex \textit{student} network to mimic the predictions of a large, pre-trained \textit{teacher} network by penalizing it for diverging from the teacher's scores~\cite{nips2014distill}.
The \textit{teacher} network is the 5-layer bidirectional GRU, shown in Table \ref{tab:ds2}, while the student models are a 4-layer GRU and a 5-layer RNN. 
Using distillation alone is insufficient to recover the baseline accuracy of the teacher network.
Instead we start with distillation and then fine-tune for ASR with CTC \cite{ctc} for 5 epochs. 
This combination yields \textit{student} networks with the same accuracy as the teacher (Figure~\ref{fig:distill}).
Compared to the 5-layer \textit{teacher} GRU and a 7-layer RNN (iso-accuracy using traditional supervised learning), the distilled 5-layer RNN has 3$\times$ and 1.4$\times$ fewer parameters, respectively.
(All recurrent layers have 800 hidden units, wherein weight matrices are 800x800.)
Knowledge distillation---reducing a dense DNNs to smaller ones ---improves performance and energy significantly and provides immediate benefits on CPUs, GPUs, and specialized hardware.

\textbf{Language Modeling} is a post-processing step that reduces the WER by modifying the output of the RNNs (after all layers and timesteps) based on language semantics and structure~\cite{ctc}.
This can be done greedily or using beam search, which maintains many likely speech-to-text transcriptions (determined by the beam-width). 
Figure \ref{fig:distill}(right) shows the decrease in WER as we increase the beam width from 1 (greedy) to 512 in increments of power of two. While the execution time for previous generations of ASR models has been dominated by the language model \cite{unfold}, this is not the case for newer ones like DS2.
With a beam width of 128, only 10\% of the CPU time is spent on performing beam search; the rest is spent running the RNN.
Furthermore, previous work has proposed specialize hardware to accelerate beam search by at least 5$\times$ \cite{darksidednnpruning}. 
Thus, for the purposes of this study, we focus on accelerating the core RNN layers, the main performance bottleneck.

 \textbf{Weight Pruning} eliminates less important weights and transforms dense matrix-vector multiplications to sparse ones.
Pruning is performed by iteratively zeroing and masking out parameters, based on absolute value, and then retraining the network~\cite{iclr2016deepcompression, reagen2017weightless, narang2017exploring}.
The number of non-zero parameters can be reduced to 33\% in the already distilled RNNs (iso-accuracy).

 \textbf{Sparse Linear Quantization} reduces the storage overheads of parameters by transforming them from 32-bit floating point type to reduced precision.
By applying simple linear quantization \cite{minerva, eie,fpga2017esehan, scnn}, the pruned and distilled network can be represented in fixed-point format with 14 bits.
However, after pruning, the remaining non-zero parameters follow a skewed distribution with either
high-negative or high-positive magnitude.
Thus, before applying quantization, we separately scale the magnitude of the positive and negative weights to fit within the range $[0,1]$.
This enables further reducing the precision down to 10 bits without sacrificing accuracy.

\input{tables/modelopt.tex}

Together, the above-mentioned optimizations improve performance, area, and energy by 4.2$\times$, 3$\times$, and 8$\times$, respectively. 
The parameters for the efficient baseline are shown in Table~\ref{tab:modelopt}.

\subsection{Supporting recurrent networks more generally} 
While the remainder of this paper focuses on accelerating the ASR RNN baseline, the key contributions of \NAME apply to recurrent neural networks with weight and activation sparsity more generally. 
First, RNNs have crafted various speech recognition networks, notably transducer (e.g., DS2), seq2seq, and attention based architectures.
Previous work has trained these architectures with ReLU activated RNNs, enabling the activation sparsity that \NAME exploits~\cite{battenberg2017exploring}. 
Next, the DS2-style RNN studied in this paper forms the encoder in multi-stage ASR networks~\cite{battenberg2017exploring,he2018streaming}.
Finally, the core micro-architectural contributions also apply to GRU-based networks with pruned weights and ReLU non-linearity for sparse activations.
Such ReLU activated GRUs have also been used in transducer, seq2seq, and attention based speech recognition networks~\cite{ds2,battenberg2017exploring,chiu2017monotonic}.

In order to optimize this vast design space of recurrent neural networks, \NAME can be configured with a combination of dynamic and design-time parameters.
Dynamic, run-time parameters include number of hidden-units, number of timesteps, and whether the RNN is uni-directional or bi-directional (see Section~\ref{sec:arch} for details).
Design-time parameters include whether the network is an RNN or GRU, and the maximum recurrent network size supported.

%% file: tables/modelopt.tex
\begin{table}[h!]
\small
\caption{\textbf{Efficient RNN baseline} after model optimizations.} \vspace{-1.0em}
  \begin{center}
    \begin{tabular}{c@{\TABSEPsmall}| c@{\TABSEPsmall}| c@{\TABSEPsmall}| c@{\TABSEPsmall}| c@{\TABSEPsmall} | c@{\TABSEPsmall}}
     \toprule
     Layer Type & Activation &  Layers & Params & Bitwidth & NZ \% \\ \cmidrule{1-6}
     Bi-dir RNN     &  ReLU & 5      &  13M & 10 & 33 \\ \hline

   \end{tabular}
\end{center}
    \label{tab:modelopt}
\end{table}

%% file: optimizations.tex
\section{Optimizing Dynamic Activations} \label{sec:opt}
As shown in Figure~\ref{fig:models}, activations are the primary memory bottleneck of RNNs. 
This section presents the methods used to enforce sparsity in activations and 
the proposed sparse encoding algorithm.

\subsection{Activation sparsification}

\textbf{Sequential processing}:
The core computation kernels of RNNs are the matrix-vector multiplications for the input and hidden states, $h^t = ReLU(W_xx^t + W_hh^{t-1})$.
These kernels can be computed either in parallel or sequentially in time.
E-PUR~\cite{epur} proposes  maximizing weight locality by first computing the input connections, $W_xx^t$, for all time steps in parallel, followed by the recurrent connections.  
Even with aggressive 10-bit quantization,
this approach requires significant on-chip storage (1.25MB), outweighing the benefits of reducing on-chip storage through weight reuse. 


\NAME\ computes each time step sequentially.
Sequential processing halves the amount of intermediate values to store. 
More importantly, as we use a ReLU activation function,
by sequentially processing each time step intermediates are sparse and amenable to compression.

\textbf{Hidden state sparsity} 
To further reduce on-chip storage requirements for activations,
\NAME makes use of the sparsity in inputs and hidden states.
Recall that our efficient baseline model is a 5-layer RNN with ReLU, i.e., $\max\{0, x\}$. 
Training with ReLU causes 80\% of the hidden state values to be zero.

\begin{figure}[t]
    \centering
    \begin{subfigure}[b]{\columnwidth}
        \includegraphics[width=0.9\linewidth]{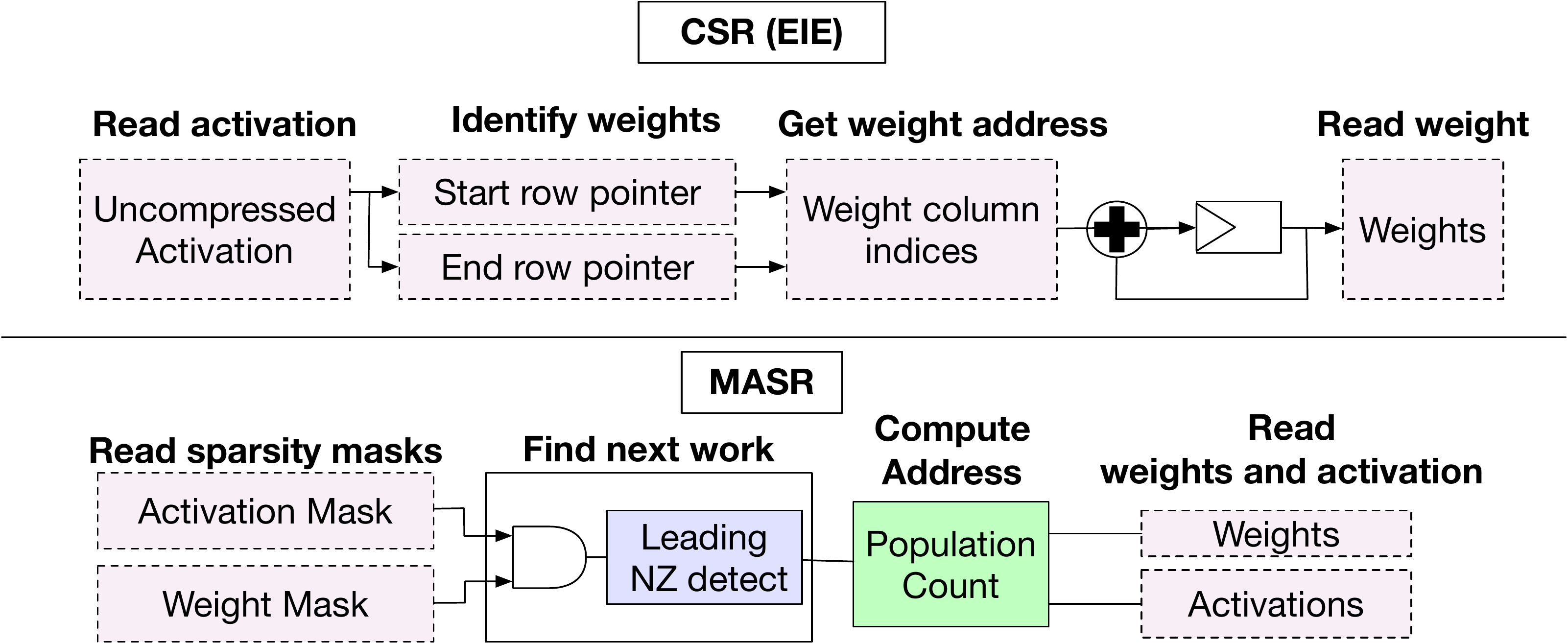}
        \caption{Block diagram of compressed sparse row encoding (EIE) and \NAME's bitmask encoding. All memory blocks are in purple. Compared to CSR, memory centric sparse encoding technique, \NAME proposes using a logic centric sparse encoding technique.}
        \label{fig:encoding_a}
    \end{subfigure} \hfill

    \begin{subfigure}[b]{\columnwidth}
        \includegraphics[width=0.9\linewidth]{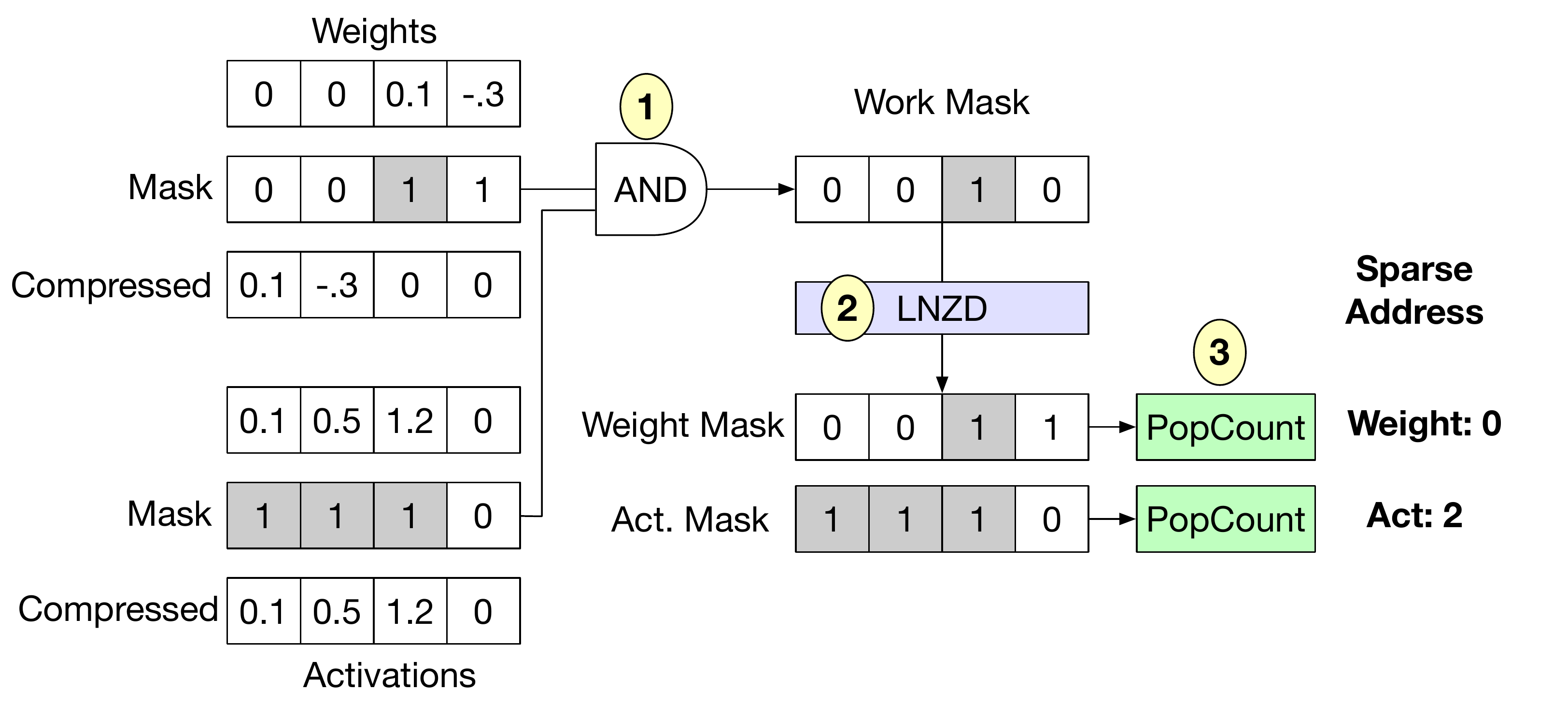}
        \caption{Concrete example of \NAME's sparse encoding. Step 1 determines the pairs of non-zero weights and activations, to produce the work mask, using a logical \texttt{AND}. Step 2 computes the next non-zero weight and activation to fetch from using with a leading non-zero detect. Finally, step 3 evaluates the address of sparse weights and activations stored compactly in memory.}
        \label{fig:encoding_b}
    \end{subfigure}

    \caption{\NAME's sparse encoding compute sparse addresses for weights and activations in logic using low-cost hardware. }
    \label{fig:encoding}%
    \vspace{-1em}
\end{figure}

\textbf{Input sparsity} 
Input sparsity is lost due to batch normalization, a regularization technique that makes training larger models easier, between layers. 
The operation adjusts and scales activations to have a zero mean and unit variance:
$x = \frac{x_{sp} - \mu}{\sqrt{\sigma^2 + \epsilon}} \times \gamma + \beta $. 
where $x_{sp}$ represents the sparse inputs and $\mu$, $\sigma$,$\epsilon$, $\gamma$, and $\beta$ represent learned parameters. 
The linear transform employs non-zero shifts (i.e., $\mu$, $\beta$) that map sparse inputs to dense ones. 
However, during inference, the linear transformation can be statically refactored into the next layer's weights at zero cost: 
\begin{align*}
x & = K_0 x_{sp} + K_1  & K_0 &= \frac{\gamma}{\sqrt{\sigma^2 + \epsilon}} & & K_1 &= \beta - \frac{\gamma \mu}{\sqrt{\sigma^2 + \epsilon}}
\end{align*}

\noindent
To refactor this computation, we multiply the next layer's weights and biases by the  $K_0$ and $K_1$ constants, respectively. 
Note that this refactoring is applicable to a broader set of neural networks that use batch normalization through depth~\cite{battenberg2017exploring, chiu2017monotonic}. 
After refactoring, inputs are on average 60\% zeros.


\begin{figure}[!t]
\centering
\includegraphics[width=\linewidth]{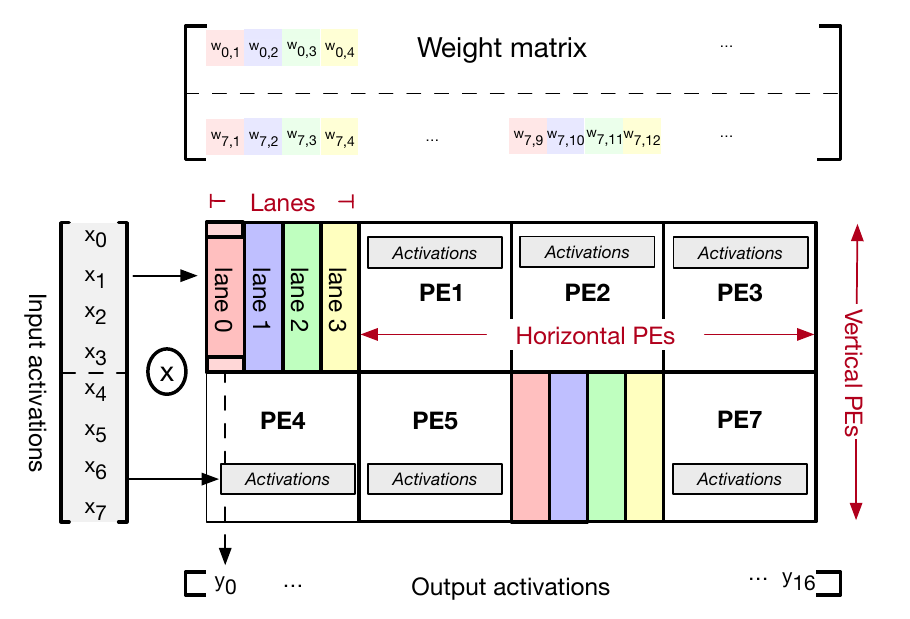}
\vspace{-1.5em}
\caption{Overall topology of how weight matrices, input activations, and output activations are split across the \NAME architecture. 
\NAME is organized as a 2D-array of horizontal (output neuron dimension) and vertical (input neuron dimension) PEs/lanes. }
\label{fig:topology}
\vspace{-1.0em}
\end{figure}

\subsection{Compact activation storage} \label{sec:sparsenc}
Operating over compressed weights and activations introduces two  challenges: (1) aligning pairs of non-zero weights and activations, and (2) generating addresses for weights and activations stored compactly in memory. 
\NAME addresses these challenges by co-designing a sparse encoding technique for both activations and weights.
As shown in Figure \ref{fig:encoding_a}, the sparse encoding technique uses a combination of bitmasks, a leading non-zero detects (LNZD), and population counts. 
We start by reading the weight and activation bitmasks. 
The bitmasks track the sparsity pattern as bit vectors, where non-zero entries are represented as ones.
Next, a bitwise \texttt{AND} between the weight and activation masks, determines pairs of non-zero weights and activations and produces the work mask. 
The ones in the work mask denote the absolute minimum work to compute.
A LNZD over the work mask determines the index of the next non-zero weight and activation to fetch from memory.
Finally, population counts of the weight and activation masks, up to the index specified by the LNZD, evaluates addresses of sparse weight and activations stored compactly in memory. 


\begin{figure*}[!ht]
\centering
\includegraphics[width=\linewidth]{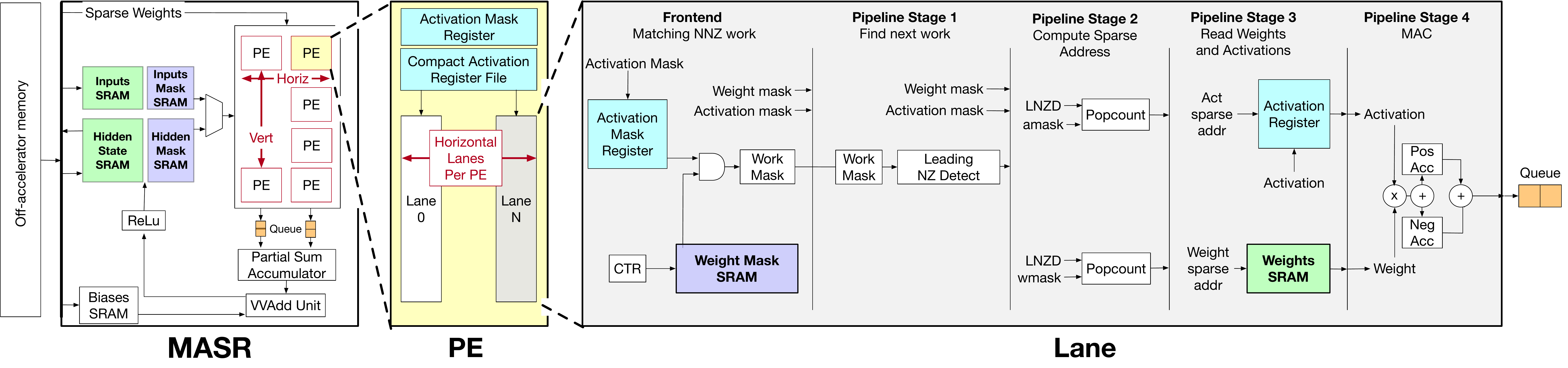}
\caption{\NAME\ accelerator design highlighting the overall system architecture (left), a PE (center), and a lane (right). Blocks outlined in red represent tunable micro-architectural parameters swept in the design space exploration. 
Also in color: activation registers (blue), and SRAMs for binary masks (purple) and for compressed sparse activations and weights (green).}
\label{fig:arch}
\vspace{-1em}
\end{figure*}

\textbf{Example \NAME encoding} Figure \ref{fig:encoding_b} provides a concrete example of \NAME's sparse encoding.
The logical \texttt{AND} between the weight ($0011$) and activation ($1110$) masks produces the work mask ($0010$).
The LNZD over the work mask points to index 2.
Population counts up to index 2 for the weight ($\textbf{00}11$) and activation ($\textbf{11}10$) masks, compute the weight ($0$) and activation ($2$) addresses, respectively.



\textbf{Comparing \NAME to run-length and CSR} 
The optimal encoding is application specific and depends on sparsity and matrix size.
Previous sparse DNN accelerators typically use run-length encoding or CSR~\cite{eie, fpga2017esehan, cambricon, scnn}. 

Run-length encoding maintains a step index that stores the distance between non-zero weights \cite{cambricon}.
However, it does not design for sparsity in activations, leaving key storage, performance, and energy savings on the table.

CSR considers sparsity in both weights and activations.
As shown in Figure \ref{fig:encoding_a}, CSR first reads the non-zero activation address, encoded using a run-length style step index. 
The non-zero activation address then indexes separate row pointer memories to identify the first and last non-zero weights corresponding to the given input activation. 
Finally, the row pointers are used to read column indices, also encoded using a run-length style step index, which generate the address of weights stored compactly. 
While this approach works well for model with high sparsity, it suffers from two main drawbacks.
First, null activations are skipped in execution not storage.
Second, while column pointers scale with the number of non-zero weights, each \textit{MAC}/PE maintains its own set of row pointers in CSR. 
As a result, row pointer memory scales with the number parallel \textit{MAC}s/PEs.
Figure \ref{fig:sparse_enc_motiv} shows that as the hardware scales from 32 to 512 parallel \textit{MAC}s/PEs, row pointers dominate the memory footprint.

\textbf{Low overhead and scalable sparse encoding} 
In contrast, the memory footprint for \NAME's sparse encoding technique does not scale with the number of parallel \textit{MAC}s/PEs.
This is a result of eliminating the row pointers and identifying the necessary sparse weights and activations by computing the alignment in logic.
For instance, \NAME computes the address of non-zero weight and activation pairs in logic, as shown in Figure \ref{fig:encoding_a}. 
The memory overheads for encoding sparsity in \NAME are limited to binary masks, which are determined by the size neural network model and \textit{not} the number of \textit{MACs}/PEs. 
Thus, \NAME has a significantly lower memory footprint compared to previous sparse neural network accelerators (i.e., EIE, ESE \cite{eie, fpga2017esehan}); see Section~\ref{sec:discuss} for a detailed quantitative comparison.


%
%

%% file: arch.tex
\section{The \NAME Architecture}\label{sec:arch}

As shown in Figure~\ref{fig:topology}, \NAME is composed of a 2D-array of processing elements (PEs)/lanes that evenly split each weight matrix in the horizontal (i.e., output neurons) and vertical (i.e., input neurons) dimensions.
Each PE is a collection of lanes that share a local activation register file.
Each lane has its own local weight and weight mask SRAMs that store an equal portion of the matrix.
Compact weight matrices (only non-zero elements) are loaded from off-accelerator memories directly into local SRAMs.
Output neurons are computed by accumulating the partial products across lanes in vertical PEs (i.e., lane 0 in PE0 and P4 determine $y_0$ in Figure~\ref{fig:topology}).
Decoupling the execution across the 32 to 1024 lanes is crucial for extracting parallelism of the irregular sparse computations at scale.
Figure \ref{fig:arch} shows the detailed architecture for \NAME, focusing on RNN computations outlined in Figure~\ref{fig:rnn}.
The modular design is centered around the 2D array of PEs and decoupled, pipelined lanes.
In this section, we first explain how bidirectional RNN computations are mapped to \NAME.
Then we show how the underlying lane micro-architecture handles sparsity in weights and activations.

\begin{figure*}[ht]
    \centering
    \includegraphics[width=\textwidth]{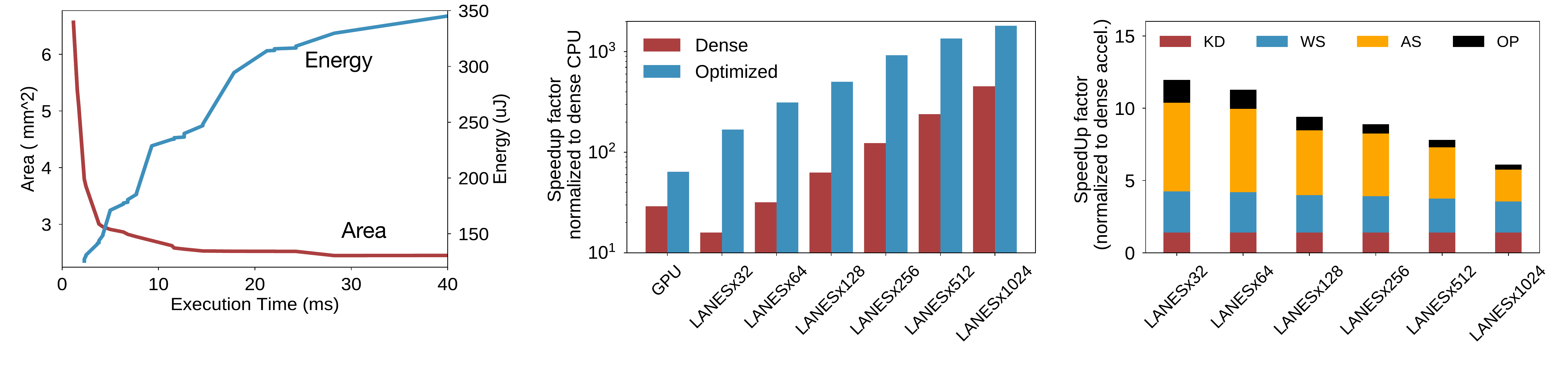}
    \vspace{-2em}
    \caption{\textbf{Left}: Energy-performance and area-performance Pareto frontiers of accelerator designs, sweeping microarchitectural parameters shown in Figure \ref{fig:arch}. \textbf{Center}: Speedup of varying \NAME designs normalized to a CPU running a dense baseline RNN. \textbf{Right}: Breakdown of performance benefits for each proposed optimization on \NAME designs.}%
    \label{fig:pareto}%
    \vspace{-1.5em}
\end{figure*}

\begin{figure}[t]
\centering
\vspace{-0.8em}
\includegraphics[width=0.90\columnwidth]{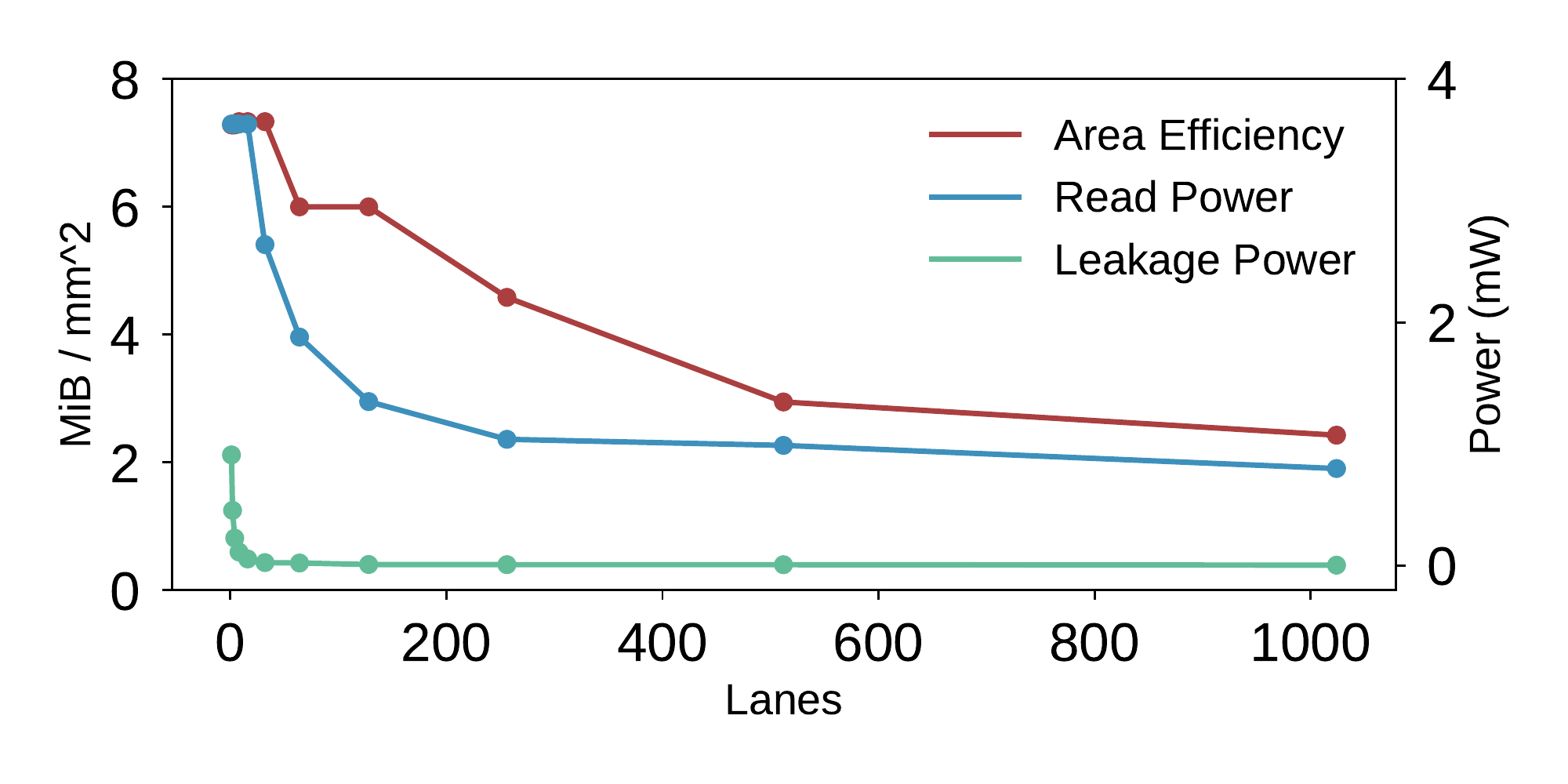}
\vspace{-1em}
\caption{Impact of increasing parallelism (lanes) on SRAM area efficiency, dynamic read power, and leakage power.}
\label{fig:sram}
\vspace{-1em}
\end{figure}

\subsection{Mapping RNN computations to \NAME}
To process speech samples, the accelerator runs each layer of the bidirectional RNN in order.
Within each layer, the accelerator first executes all time steps in the forward direction and then in the backward direction.
Recall that each time step of the RNN comprises two matrix-vector multiplications, a vector-vector addition and ReLU: $h^t = \textrm{ReLU}(W_x x^t + W_h h^{t-1})$.
To begin processing a layer, all weights for the forward pass ($W_x$ and $W_h$) are loaded from off-accelerator memory into the compact weight SRAMs within the lanes. Weight SRAMs have a word width of 10 bits (1 weight each).
Likewise, all inputs ($x^t$) are loaded into compact activation SRAMs, which have a word width of 60 bits (six activations each).

While \NAME processes all time steps in the forward pass, weights for the backward pass are concurrently loaded into separate SRAMs.
This double buffering of the forward and backward weights reduces performance penalties from not having the entire layer's weights stored locally on-chip.
Similarly, activations beyond 333 timesteps (the average length of speech samples in Librispeech) are also double buffered.
Section \ref{sec:resource} discusses the design decisions of double buffering.

{\bf Hidden state computation:}
For each time step $t$, the accelerator first processes the matrix-vector multiplication for hidden state, $W_hh^{t-1}$.
This computation is initiated by loading the previous time step's hidden states from the compact activation SRAM to compact activation register files within each PE.
The entirety of the matrix-vector multiplication is parallelized across the 2D array of PEs with multiple decoupled lanes.
Horizontal lanes evenly split columns (output dimension) of the matrix, while vertical lanes evenly split rows (input dimension).
This enables balancing parallelism across the input and output dimensions.
Each lane is responsible for computing partial products for a subset of rows and columns in the matrix-vector product.
As lanes finish processing each column, the partial sum accumulator sums the partial products for each output.
An 800 element register file stores the outputs.

{\bf Finishing one time step:} The above sequence repeats to process the matrix-vector multiplication for inputs, $W_x x^t$.
Once both matrix-vector multiplications have been processed, the vector-vector add unit accumulates the biases, input intermediates, and hidden intermediates.
The resulting output values are thresholded with ReLU and compactly written to the hidden-state SRAM for the subsequent time step.
This completes one time step of the RNN layer.

\begin{figure*}[t]
    \centering
    \includegraphics[width=0.99\linewidth]{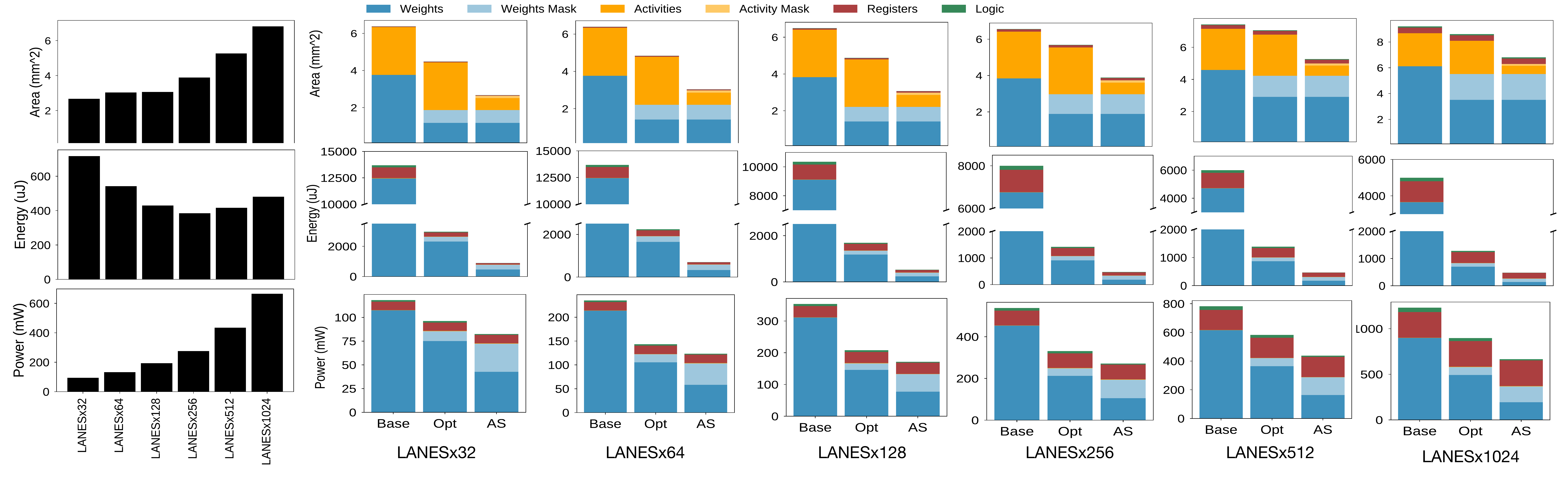}
    \caption{ The plots on the left summarizes area (top row), energy (middle row), and power (bottom row) tradeoffs for the fully optimized designs for various \NAME design points.
    On the right we breakdown each optimization and each resource (weights, activations, sparse encoding masks, registers, and logic).
    The breakdowns are for the optimized baseline (distilled, Section~\ref{sec:modelopt}), optimized (weight pruning), and Act (sparse activations, fully optimized).
    }
    \label{fig:ppa}%
    \vspace{-1.5em}
\end{figure*}

{\bf Hardware implications of parallelism:}
The number of parallel lanes determines the degree of parallelism and how weights are partitioned across SRAMs.
With 1024 lanes, each lane's weight and weight mask SRAM stores $\frac{1}{1024}$ of the parameters.
Similarly, number of vertical lanes determines the size of the compact activation register files.
For example, with 32 vertical lanes, each register file only tracks $\frac{1}{32}$ of the values.
Horizontal lanes in a row process the same portion of the activation vectors.
To reduce the cost of duplicated activations, horizontal lanes within a PE share a physical activation register file.
Given that lanes are decoupled, increasing the number of lanes per PE requires additional ports to the physical register file, which increases the register file's size and cost per access.
Note that \NAME's decoupled PE/lane architecture does not depend on complex crossbar architectures that can limit efficiency of highly parallel sparse DNN accelerators.
Section \ref{sec:dse} explores the design space encompassed by these parameters.
Table \ref{tab:designs} illustrates the parameters for LANESx32, LANESx256, and LANESx1024.

Outside of the PEs, the \NAME architecture has two additional parameters: depth of the back end queues and number of banks for activation SRAMs.
The partial sum accumulators accumulate the output of each column once all lanes in the given vertical slice finish generating their partial product.
Lanes that finish early are stalled, reducing the performance of the overall design.
Increasing the depth of the back end queues reduces this back pressure.
However, this comes at an area and energy cost, given each lane pushes partial products to a separate back end queue.
In addition, the vector-vector add unit can be parallelized.
This involves not only duplicating the number of adders but also partitioning the activation SRAMs into multiple banks (see Section \ref{sec:inefficiency} for details).

\input{tables/tab_designs.tex}
\subsection{The \NAME\ Lane}
The \NAME\ lane is the main computational workhorse for sparse vector-matrix multiplication.
Each pipelined lane is organized in two phases, front end and back end.
Intuitively, the front end decodes work from weight and activation masks, whereas the back end performs \textit{MAC}s after accessing compact weights and activations.
This eliminates wasted work in the back end, regardless of the distribution of sparse weights, activations, and outputs.

The front end first reads the binary masks, for both weights and activations.
The binary masks are then \texttt{ANDED} together and the resulting \textit{work} mask represents the absolute minimum non-zero weights and activations to accumulate.

The back end has four pipeline stages.
Stage~1 receives the work mask and uses a single-cycle LNZD to find the next pair of non-zero weights and activations.
Stage~2 computes the relative addresses using the LNZD output and population counts of the weight and activation masks.
Stage~3 reads the weights SRAM and activation register file.
Stage~4 evaluates the \textit{MAC}.
Separate accumulators are maintained for the positive and negative weights as they were quantized separately (see in Section 2).
When the computation for the output neuron finishes, the partial sum is pushed onto the queue.

\section{Evaluation Methodology} \label{sec:eval}
The accelerator design space we explore is vast and each point is evaluated running the entire forward and backward passes of the bidirectional RNN.
We validate a custom cycle-level C++ simulator of the accelerator with a synthesized RTL implementation.
We  annotate the simulator based on PPA characterizations from synthesized RTL using a commercial 16nm FinFET standard cell library at 1GHz.
To model the SRAM area, energy, and power consumption, we use a commercial memory compiler in the same process.


We also evaluate the benefits of sparsity on CPUs and GPUs by profiling GEMM and SPMV kernels on real machines.
For CPU baselines we run the Eigen library on a desktop Intel Core i7-6700K with SIMD support, using the \texttt{-O3} and \texttt{-ffast-math} compiler flags.
The GPU baselines run GEMM and SPMV kernels provided by Deep Bench~\cite{deepbench}, using cuBLAS/cuSparse libraries on a NVIDIA GTX 1080 GPU.

\input{tables/pareto.tex}

%% file: tables/tab_designs.tex
\begin{table}[t]
\small
  \begin{center}
    \caption{
     \NAME design parameters}
    \label{tab:designs}
    \begin{tabular}{c|c|c|c}
     \hline
       Lanes          & 32 & 256 & 1024  \\  \hline
      Weights per lane (KB)  & 32 & 4 & 1 \\ \hline
      Weight masks per lane (KB) & 10 & 1.25 & 0.3 \\ \hline
      Total weight (KB) & \multicolumn{3}{c}{1280} \\ \hline
      Total activations (KB) & \multicolumn{3}{c}{450} \\ \hline
      Weight width (bits)  & \multicolumn{3}{c}{10}  \\ \hline
      Activation width (bits) & \multicolumn{3}{c}{10} \\ \hline
      Technology node & \multicolumn{3}{c}{16nm}  \\ \hline
      Frequency (MHz) & \multicolumn{3}{c}{1000} \\ \hline
   \end{tabular}
  \end{center}
  \vspace{-2.5em}
\end{table}

%% file: tables/pareto.tex
\begin{table}[t]
\small
  \begin{center}
    \caption{
     Topology of \NAME\ Pareto front points.
}
    \label{tab:pareto}
    \begin{tabular}{c|c|c|c}
     \toprule
      Accel & Horiz Lanes & Vert Lanes & Horiz PEs  \\  \hline
      LANESx32 & 16 & 2 & 2 \\ \hline
      LANESx64 & 32 & 2 & 2 \\ \hline
      LANESx128 & 32 & 4 & 2 \\ \hline
      LANESx256 & 32 & 8 & 2 \\ \hline
      LANESx512 & 32 & 16 & 1 \\ \hline
      LANESx1024 & 32 & 32 & 1 \\ \hline
   \end{tabular}
  \end{center}
  \vspace{-2em}
\end{table}

%% file: designspace.tex
\section{Perf., Area, Energy, and Power Benefits} \label{sec:dse}
Optimal configuration of \NAME's modular architecture depends on the intended use case.
This section presents results of an extensive design space exploration of \NAME's free parameters that exposes energy-performance tradeoffs.
We then analyze the performance, area, and energy/power breakdowns for points along the Pareto frontier of the design space in order to quantify the benefits of each optimization and identify where resources are being consumed.
For all experiments in this section, we fix the depth of the back end accumulator queue to a single element and assume one-bank activation SRAMs.
We report the performance, area, and energy consumed for an accelerator provisioned to run a full seven seconds of speech, the average sample length in the Librispeech corpus, across multi-layer bidirectional RNNs.
Finally, we discuss how the design scales when running shorter and longer speech samples.

\subsection{Design Space Exploration}

\NAME's modularity enables both high performance and low power solutions.
The tunable microarchitectural parameters considered in the design space, outlined in Figure~\ref{fig:arch}, include the number of horizontal lanes, number of vertical lanes, and number of horizontal PEs.
All possible configurations are swept so that the total number of lanes ranges from 1 to 1024 at powers of 2, with a maximum of 32 lanes in either dimension.

Sweeping the total number of lanes produces the energy-area-performance Pareto frontiers illustrated in Figure~\ref{fig:pareto} (left).
As parallelism increases, execution time and energy consumption decrease while area increases.
This is a result of partitioning SRAMs into smaller arrays in order to support the bandwidth needed for more parallel datapaths. 
Figure~\ref{fig:sram} shows that partitioning SRAMs decreases the power per read and per-bit area efficiency.
Even in highly parallel architectures such as LANESx1024, SRAM leakage is a small fraction of the overall energy due to the highly optimized 16nm FinFET libraries.

In addition to lane count the organization of lanes/PEs has an impact on accelerator performance. 
Table~\ref{tab:pareto} shows Pareto optimal designs tend to have more horizontal lanes than vertical ones. 
Increasing the number of vertical lanes reduces the number of rows each lane processes, and thus also reduces the activation register file size. 
For example, with 8 vertical lanes, the activation register files contain 64 words; with 32 vertical lanes, they contain 16. 
Processing a small number of activations leads to load imbalance across lanes, degrading performance. 
A solution to this problem is discussed in Section \ref{sec:inefficiency}.
The following sections detail the performance, area, energy, and power characteristics of the optimal designs.


\subsection{Performance}
Figure \ref{fig:pareto} (center) shows the speedup by running a dense 7-layer bidirectional RNN and the efficient baseline on \NAME and a GPU, normalized to running the dense network on a CPU.
Performance is measured as the execution time to process the full 7-seconds of speech.
While the more programmable GPU benefits from knowledge distillation (1.4$\times$) and weight pruning (3$\times$), it is unable to exploit activation sparsity.
The \NAME designs benefit from knowledge distillation, weight pruning, and sparse activation execution, improving the performance of the accelerator beyond that of the more programmable systems.

Figure~\ref{fig:pareto} (right) breaks down the performance benefit that each optimization offers. 
We find that the overall benefits of sparse optimizations diminish as parallelism increases.
While each design observes speedup from knowledge distillation and sparse weight execution, speedup from sparse activation execution does not scale as gracefully.
For example, we find sparse activations provide up to 3.75$\times$ speedup on the LANESx32 design, while their benefit on LANESx1024 is reduced to 2.2$\times$ due to higher load imbalance in more parallel accelerator topologies.
Section~\ref{sec:inefficiency} proposes low-cost solutions to balance dynamic activations at run-time --- improving speedup from sparse execution by 1.7$\times$ and allowing even the most parallel designs to achieve near-linear speedup.

\input{predicated_neuron.tex}

 \begin{figure}[t]
    \centering
        \includegraphics[width=\linewidth]{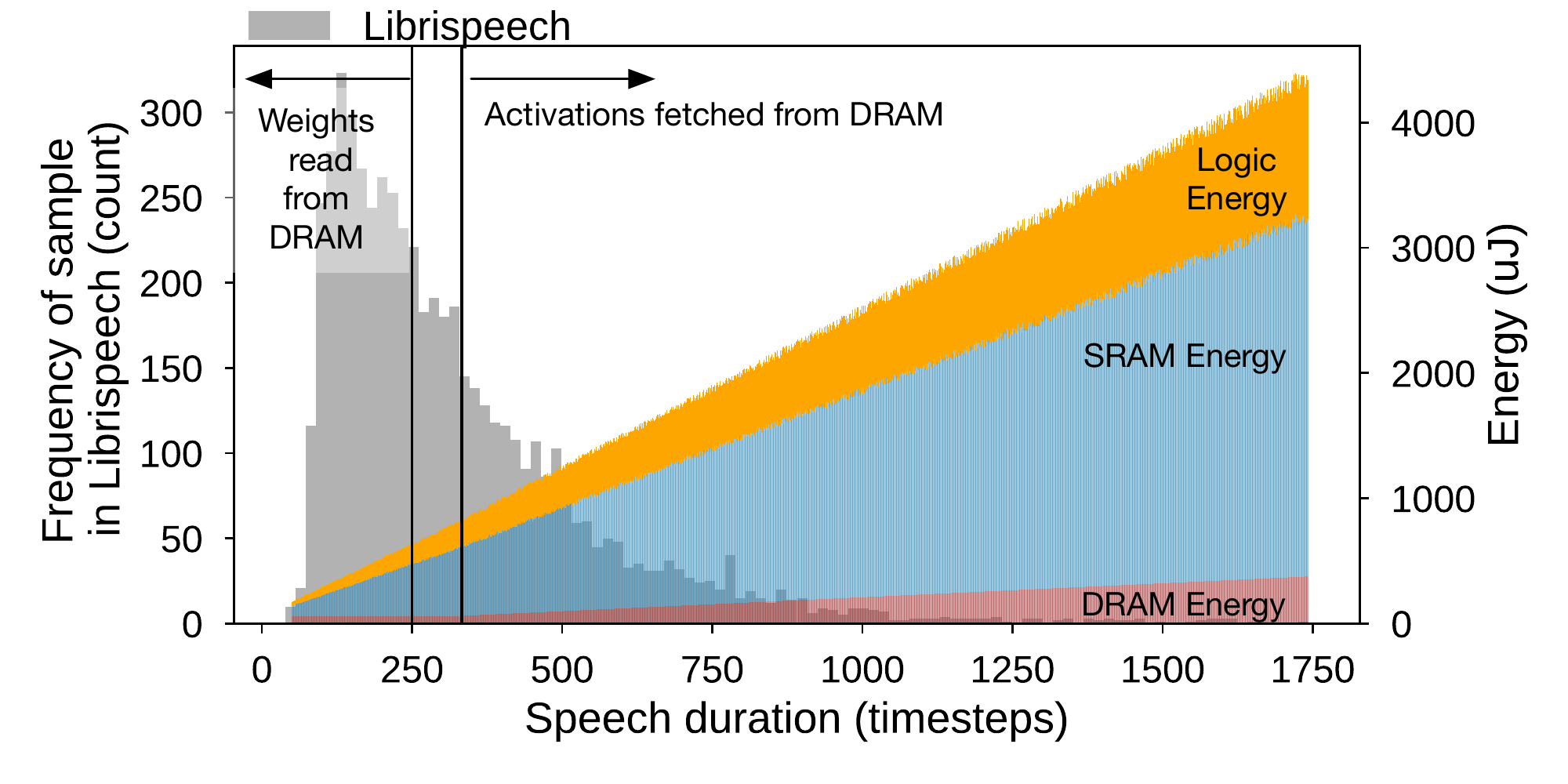}%
\caption{ Energy breakdown of LANESx256 running speech of varying length overlaid with the distribution of samples found in Librispeech. 
DRAM energy cost, for double buffering weights and activations, is small.}
    \label{fig:energy_timesteps}%
    \vspace{-1.0em}
\end{figure}


\begin{figure*}[ht]
    \centering
    \includegraphics[width=\textwidth]{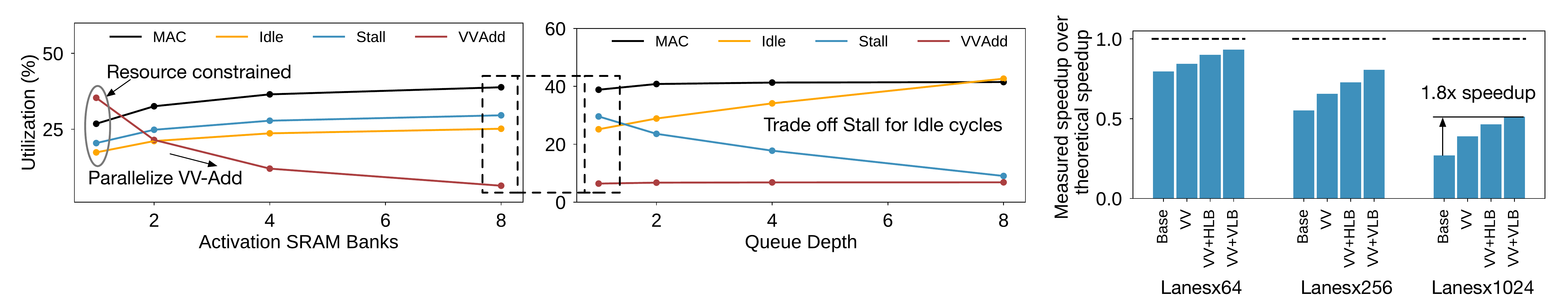}
    \caption{\textbf{Left}: As we scale the number of activation SRAM banks in the LANESx1024 architecture, the cycles spent on VVAdd decreases.
    \textbf{Center}: Scaling the depth for back-end accumulator queue trades off stalls for idle cycles.
    \textbf{Right}: The impact of increasing the number of activation SRAM banks (VV) combined with either horizontal load balancing (VV+HLB) or vertical load balancing (VV+VLB) on the performances of LANESx64, LANESx256, and LANESx1024..}%
    \label{fig:utils}%
    \vspace{-1em}
\end{figure*}

\subsection{Area, Energy, and Power} \label{sec:resource}
Figure \ref{fig:ppa} shows the area, energy, and power breakdowns for each design point along the energy-performance Pareto frontier.
The left column illustrates the overall trends as the accelerator scales to more parallel design points. 
The remaining columns on the right side provide detailed resource breakdowns.
To understand the benefits of each optimization in accelerators with varying degrees of parallelism, three sets of resource breakdown bars correspond to design variants provisioned to run a dense 7-layer bidirectional RNN (\textit{Base}), the efficient baseline that applies knowledge distillation and weight pruning (\textit{Opt}), and with sparse activations (\textit{AS}). 


\textbf{Area:} Figure~\ref{fig:ppa} (top row, right) shows the area footprints of each accelerator design. 
Partitioning weight and weight mask SRAMs diminishes the benefits of compactly storing weights in more parallel designs.
For instance, for the LANESx64 architecture, starting from (\textit{base}), weight sparsity (\textit{opt}) reduces area from 5.5$mm^2$ down to 4.0$mm^2$ (1.5$mm^2$ benefit).
On the other hand, weight sparsity reduces LANESx1024 area from 8.3$mm^2$ to 7.8$mm^2$ (0.5$mm^2$ benefit).

After compressing the weights, quantized activations consume up to 50\% of the accelerator area, especially in the smaller LANESx32-256 designs.
Compact activation storage reduces the memory area consumed by activations by 3$\times$.
This corresponds to reducing the area devoted to activations alone from 1.7$mm^2$ to 0.6$mm^2$.
Given \NAME's modular design, compact activation storage provides the same area benefit to all design variants; input and hidden-state memories are maintained outside of the PEs/lanes.
This modularity also facilitates scaling the architecture to domains that may require processing much larger speech samples~\cite{duplex}. 
For instance, provisioning \NAME to process up to 15 seconds of speech, compact activation storage would save 2.2$mm^2$, reducing overall area by 1.8$\times$.

After weights and activations, the remaining area is consumed by registers and logic.
The increase in register area across more parallel designs is dominated tracking more weights and activation bitmasks per lane.
These bitmasks account for over 90\% of the register area.
The secondary consumers (8\%) of the register area are the backend queues.

\textbf{Energy and Power:} The energy breakdown across each accelerator design is shown Figure~\ref{fig:ppa}(middle row).
The left column shows that LANESx256 is the energy-optimal design point even though LANESx1024 uses smaller SRAMs that dissipate less read power.
The reason is two-fold:
per-read energy cost plateaus in the most parallel accelerator designs, and the proportion of power consumed by registers increases for larger accelerators.
As previously discussed, the first effect occurs because the smaller memories used in the more parallel designs (LANESx512 and LANESx1024) do not reduce dynamic read power proportionally to capacity reduction.
The second effect comes from more parallel designs requiring the maintenance of more active states.

Generally, energy savings come from doing less work and making fewer SRAM accesses. 
For example, Figure \ref{fig:ppa} (middle) shows that compared to the dense RNN (\textit{base}), the efficient baseline reduces the energy consumed by 4.2$\times$ (i.e., 1.4$\times$ from knowledge distillation, 3$\times$ from weight pruning) across all accelerator designs.
Similarly, sparse activation execution further reduces energy by around 2.5$\times$.

Because of \NAME's sparse encoding mechanism,
sparsity optimizations impact energy more than power. 
As long as work remains, a \textit{MAC} is issued to the lane on every cycle, keeping power relatively constant.
The 1.4$\times$ power reduction between dense RNN (\textit{base}) and optimized baseline (\textit{opt}) comes from decreasing the size of weight SRAMs by storing fewer non-zero parameters (Figure~\ref{fig:ppa}, middle and bottom rows).


\subsection{Supporting Speech of Arbitrary Length} \label{sec:doublebuffer}
ASR models comprise millions of parameters which cannot be realistically stored on-chip.
The storage requirements are further exacerbated when considering activation memory for longer speech samples.
To minimize on-chip SRAM, \NAME double buffers  both weights and activations. 

\textbf{Performance and area} Using LPDDR4 as off-accelerator memory, the performance and energy penalty of double buffering is \emph{relatively} low. 
\NAME double buffers forwards and backwards weights in separate SRAMs. 
LPDDR4 supports a bandwidth of 25.6GB/s while dissipating 200$mW$ of power \cite{drampower}.
At this rate it takes 0.019ms to read in a layer's weights.
This corresponds, roughly, to the time it takes process 250 timesteps of speech. 
Thus, to avoid memory contention, \NAME stores activations for the first 333 timesteps (the average length in Librispeech) on-chip. 
Activations for later timesteps are double buffered within an 800-element register, which incurs negligible area overheads. 
Similarly, there is no performance penalty since the time to read activations from LPDDR4 is strictly less than the time to process a single timestep.

\textbf{Energy} Figure \ref{fig:energy_timesteps} illustrates the DRAM energy cost relative to on-chip SRAM and logic for samples from 50 timesteps to 1600 timesteps.
For samples less than 333 timesteps, the DRAM energy consists solely of reading weights.
This energy overhead is amortized with longer speech samples. 
For instance, at 333 timesteps, DRAM consumes about 10\% of the energy. 
Speech samples longer than 333 timesteps, incur an additional energy penalty for reading activations from DRAM; however, DRAM energy remains a small fraction compared to SRAM and logic. 
Thus, \NAME supports arbitrary length speech samples at negligible performance, area, and energy cost.

%% file: predicated_neuron.tex
\textbf{Output sparsity}
In addition to exploiting input neuron sparsity, $x^t$ and $h^t$, prior work exploits sparsity in output neurons~\cite{zhu2018sparsenn,SnaPEA}.
This requires predicting output neurons that will be masked by ReLU.
While the focus of paper is on exploiting weight and activation sparsity, \NAME can also support output sparsity.
In particular, input intermediates, $W_x x^t$, and hidden-state intermediates, $W_h h^t$, follow distinct distributions.
Batch-normalizaton only operates on inputs, $x^t$ (zero mean) causing hidden-state intermediates to be more negative than input intermediates.
Highly negative hidden intermediates, computed first, will likely be zeroed out by the ReLU function even after accumulating with input intermediates.
Thus, input intermediate calculations are skipped if the corresponding hidden intermediate is sufficiently negative, akin to the output sparsity predictor in ~\cite{zhu2018sparsenn}.
Figure \ref{fig:pareto}(right) shows that output predication (\textit{OP}) improves performance by up to 15\%.

%% file: inefficiencies.tex
\section{Scalability for end-to-end RNN} \label{sec:inefficiency}
As the number of parallel lanes increases, two main performance bottlenecks emerge: vector-vector add (VVAdd) operations and load imbalance.
We address these bottlenecks by: (1) parallelizing the VVAdd operations with multi-banked activation SRAMs; and (2) dynamic load balancing for sparse activations.
These optimizations improve performance by up to 1.8$\times$, allowing highly parallel designs to achieve high \textit{MAC} utilization while executing with sparse weights and activations.

\textbf{Parallelizing VVAdd}
Each time step in RNNs includes two matrix-vector multiplications and a VVAdd (i.e., $W_x x^t + W_h h^{t-1}$).
Although the matrix-vector multiplications are the core kernels, Figure~\ref{fig:utils}~(left) shows that with a single activation SRAM bank, the LANESx1024 design's \textit{MAC} utilization is only 27\%, while the largest fraction of cycles (35\%) devoted to VVAdd.
This is due to bandwidth limitations of the activation SRAMs.
Recall that the word width of the activation SRAM is 60 bits, limiting
VVAdd operations to 6 per cycle.
Partitioning the compact activation memory into smaller banks enables parallelizing the computation.
Partitioning the memory into 8 banks decreases the fraction of cycles spent on VVAdd operations from 35\% to 6\%, and increases the \textit{MAC} utilization from 27\% to 39\% at a negligible area penalty.

\subsection{Dynamic Activation Load Balancing}

After parallelizing the VVAdd operation, the next bottleneck to scaling performance is load imbalance, a result of irregular sparsity.
Previous work uses load balance-aware pruning, where the network is pruned during training such that each \textit{MAC} gets the same number of non-zero weights~\cite{fpga2017esehan}. This static method does not work for dynamic activation sparsity.
Moreover, the main source of load imbalance in RNNs is the uneven distribution of non-zero activations across PEs.
To address this dynamic load imbalance, we first trade off stall cycles for idle cycles by increasing the depth of back end queues.
We then propose a low-cost solution to dynamically redistribute work to idle lanes, which improves the \textit{MAC} utilization and performance of the LANESx1024 architecture by 1.3$\times$.

Figure \ref{fig:utils} (center) illustrates the trade-off between stall and idle cycles as the back end accumulator queue depth increases.
Stalls are caused by back pressure from the accumulator.
Idle cycles are caused by some lanes computing their partial outputs faster than others.
With a queue depth of one, lanes spend 28\% of their time stalled, and 22\% of their time in the idle state.
By increasing the depth of the queues to 8, the fraction of stall cycles falls to 9\% whereas the fraction of idle cycles climbs to 42\%. 0.3mm$m^2$.
This comes at a negligible area penalty of 0.3mm$^2$ for the largest LANESx1024 design.

 \begin{figure}[t]
    \centering
    \includegraphics[width=0.85\linewidth]{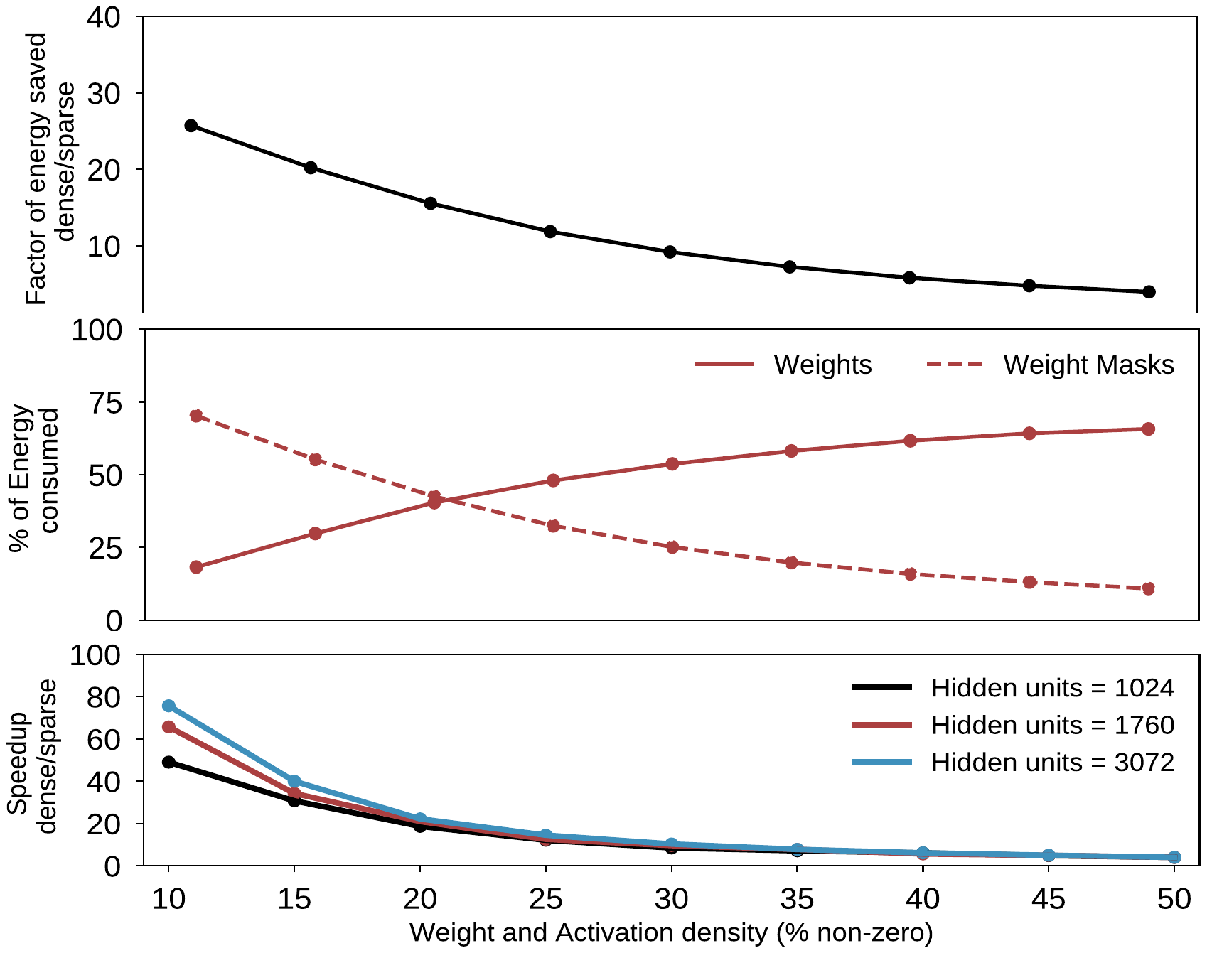}%
\caption{Energy savings (\textbf{top}), energy consumption (\textbf{middle}), and performance impact of scaling to larger RNNs (\textbf{bottom}) as density in weights and activations varies on  LANESx256.}
    \label{fig:flexibility}%
    \vspace{-1.5em}
\end{figure}

\textbf{Balancing non-zero activations} The high percentage of idle cycles suggests redistributing work to lanes that finish early, by balancing load across both horizontal lanes and vertical lanes.
With horizontal load balancing, work is distributed to lanes within the same PE.
Since all lanes within a PE process the same activations, horizontal load balancing requires storing additional copies of compact weights for neighboring lanes.
In practice, we only duplicate about 10\% of the weights.
Vertical load balancing distributes work to lanes across vertical PEs.
This involves duplicating not only weights but also activations.
Given activations are stored in local registers, the cost of duplicating them is negligible.
Duplicating weights and activations to enable load balancing also eliminates the need for complex crossbar inter-connects that often limit efficiency for sparse DNN accelerators at scale.

\textbf{Hardware utilization} Figure \ref{fig:utils} (right) illustrates the impact of each optimization on the performance of LANESx64, LANESx256, and LANESx1024.
To highlight how well each design variant parallelizes sparse RNNs, we normalize the performance of each to its theoretical speedup over serial execution.
Vertical load balancing outperforms horizontal load balancing, because redistributing work across PEs balances the number of non-zero activations, the main source of load imbalance.
Moreover, LANESx64, LANESx256, and LANESx1024 designs achieve 90\%, 80\%, and 50\% utilization.


\subsection{Scaling RNN Size and Sparsity}
Recent advances in the machine learning community allow training sparser networks without sacrificing accuracy \cite{iclr2016deepcompression,softweightsharing,dns}.
This suggests further performance and energy improvements may be possible with even higher sparsity.
To study the robustness and scalability of \NAME, we artificially scale weight and activation non-zero ratios, using synthetic RNN benchmarks, for the energy-optimal LANESx256 design.


Figure~\ref{fig:flexibility}(top, middle) plots the energy savings and consumption as the non-zero ratio in weights and activations scales from 10\% to 50\%.
As the energy saved from sparse execution depends on the non-zero ratio (not model size), we consider RNNs with 3072 hidden states.
\NAME's energy efficiency improves with greater sparsity, a result of fewer memory accesses.
For example, the energy savings at non-zero ratios of 25\% and 10\% are 12$\times$ and 26$\times$ respectively.
At lower non-zero ratios, sparse encoding overheads limit savings as weight masks dominate energy consumption.

Figure \ref{fig:flexibility}(bottom) plots the impact of sparsity on performance across a range of network sizes.
As expected, speedup from sparse execution improves with greater sparsity.
For RNNs with 3072 hidden units, sparse execution yields a 14.4$\times$ and 76$\times$ speedup with 25\% and 10\% non-zeros, respectively.
Finally, we find that performance improvements of sparse execution scale better for larger models.
For instance, with 10\% non-zeros, the speedup is 49$\times$ for the RNNs with 1024 hidden unit.
This due to better load balancing in larger models.
Thus, we expect \NAME\/'s architecture to scale well with larger ASR RNNs and advanced pruning techniques are applied.

\subsection{Hardware Implementation}
\begin{figure}[t!]
    \centering
    \includegraphics[width=0.65\linewidth]{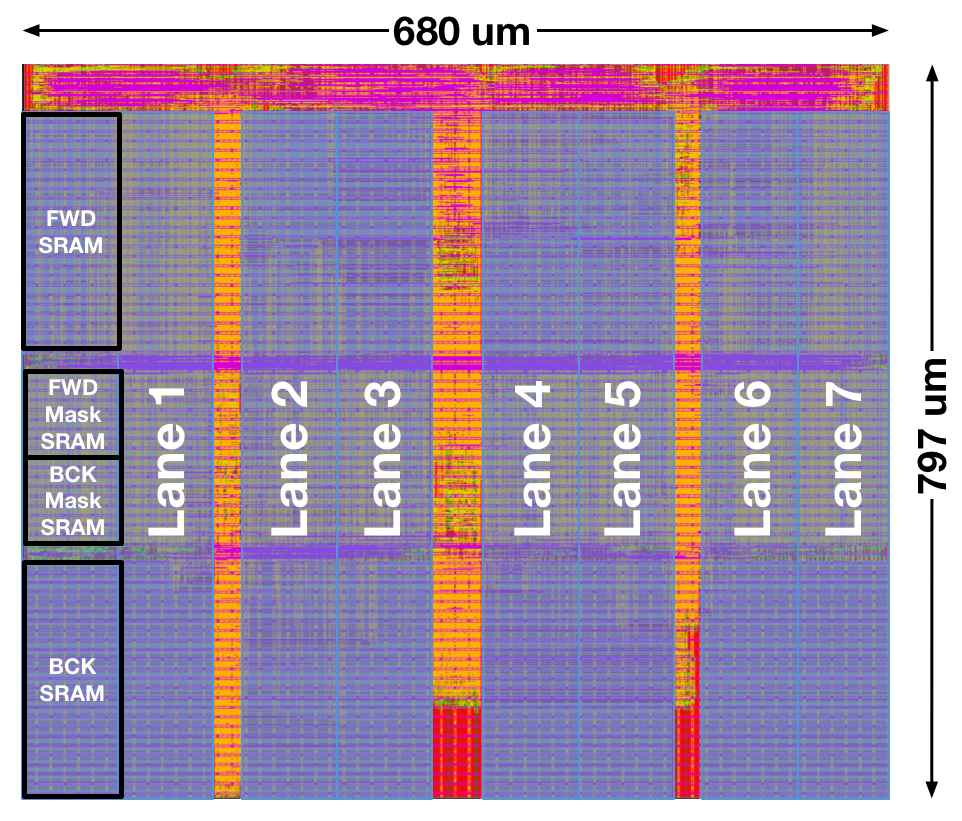}
\caption{\NAME\ LANESx32 placed-and-routed layout}
\label{fig:layout}
\vspace{-1em}
\end{figure}

Results shown thus far are based on cycle-level C++ simulations
with power models derived from synthesized RTL.
A PE of a LANESx32 RTL was placed-and-routed, as shown in Figure~\ref{fig:layout}, using a commercial 16nm FinFET standard cell library and memory compiler.
We validate our simulation results within 10\% power and 12\% area and find negligible difference in performance.
A fabricated SoC based on LANESx32 design has been received from fabrication.

%% file: discussion.tex
\section{Discussion} \label{sec:discuss}
This section provides a quantitative comparison between the \NAME accelerator and two other accelerators, shown in Table~\ref{tab:related}, for sparse neural networks with different design objectives: ESE \cite{fpga2017esehan} (weight sparsity) and EIE \cite{eie} (both weight and activation sparsity).
Based on each accelerator's memory access patterns, we accumulate the cost of the weights memory, activations memory, and sparse indexing.
For fair comparison, each design is implemented as a specialized ASIC with the same 16nm FinFET process and the same optimized RNN, see Table~\ref{tab:modelopt}.


\textbf{Performance} 
Assuming the same weight sparsity, activation sparsity, and number of parallel \textit{MACs/PEs}, hardware utilization determines performance differences between the accelerators.
The LANESx256 design for \NAME demonstrates an 80\% utilization, compared to 50\% in EIE~\cite{eie}. 
This is a result of ensuring no wasted work with the binary mask sparse encoding and re-distributing sparse activations for dynamic load balancing in \NAME.
For instance, CSR adds superfluous non-zero values (up to 40\% wasted work) and does not account for imbalance in non-zero activations.
By exploiting activation sparsity, \NAME has 3$\times$ higher performance than ESE.

\begin{figure}[t]
   \centering    \includegraphics[width=0.85\linewidth]{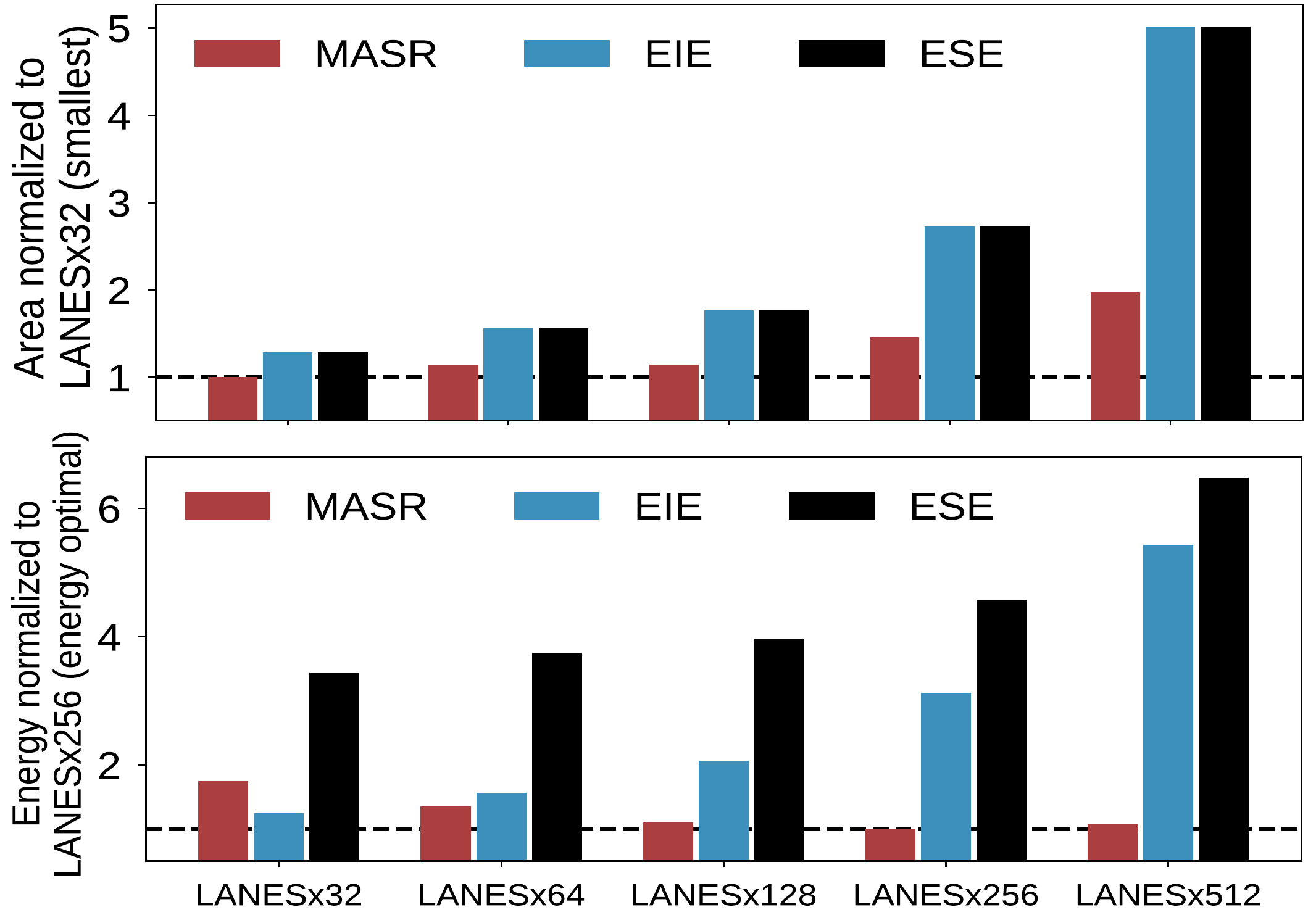}
    \caption{Area normalized to LANESx32 (top) and Energy normalized to LANESx256 (bottom) consumed by \NAME, EIE and ESE as parallel \textit{MAC}s/lanes scale. All accelerators are implemented in same 16nm process technology. }
    \label{fig:comparison}%
    \vspace{-1em}
\end{figure}

\textbf{Area} Figure \ref{fig:comparison} (top) compares the area footprints of \NAME, ESE, and EIE, all normalized to the area of the smallest \NAME design (LANESx32), as the designs scale from 32 to 512 parallel \textit{MAC}s/lanes.
The area for ESE and EIE are equivalent, as both  store activations densely and weights compactly, using CSR. 
For smaller architectures, such as those with 32 or 64 parallel \textit{MAC}s/lanes, \NAME's area savings come from compactly storing sparse activations.

Area savings are more pronounced as the accelerators scale to higher parallelism due to \NAME's lower-overhead sparse encoding mechanism. 
Storage for row pointers, used in CSR in ESE and EIE, scales  with the number of \textit{MAC}s and dominates for accelerators with more than 128 parallel \textit{MAC}s . 
Instead of explicitly storing the row pointers in memory, \NAME computes the addresses for sparse weights and activation in logic (see Section \ref{sec:sparsenc} for details).
This consumes a fixed amount of memory regardless of the number of parallel lanes.
As a result, \NAME\ has at least 2$\times$ smaller on-chip area footprint as accelerators scale beyond 128 \textit{MAC}s/lanes.
 
\textbf{Energy} Figure \ref{fig:comparison} (bottom) compares the energy consumed by \NAME, ESE, and EIE, all normalized to the energy of the energy-optimal design (LANESx256) as the designs scale from 32 to 512 parallel \textit{MAC}s/lanes.
In addition to the area benefits, \NAME\ consumes 5$\times$ and 3$\times$ less energy than ESE and EIE, respectively, as it scales beyond 128 \textit{MAC}s/lanes.
These energy savings come from \NAME's lower overhead sparse encoding mechanism.
For each row in the matrix, a PE in ESE and EIE reads two row pointers to determine the first and last non-zero weights.
Row pointer accesses scale with the number of parallel, and, like the area overheads, energy consumption is dominated by these row-pointers for more parallel designs.
\NAME\ eliminates the cost of reading row pointers by computing the sparse indexing in logic. 

%% file: conclusion.tex
\section{Conclusion}
We present \NAME, 
a novel bidirectional RNN accelerator for on-chip ASR that exploits sparsity in both dynamic activations and static weights, compacts storage of non-zero parameters, and wastes no energy at all on null computations.
Compared to a state-of-the-art sparse DNN accelerator~\cite{eie}, \NAME improves performance, area, and energy by 1.6$\times$, 3$\times$, and 2$\times$, respectively.
\NAME's modular architecture provides scalable designs ranging from resource-constrained low-power IoT applications to highly parallel datacenter deployments.

\section*{Acknowledgements}
This work was supported by the Applications Driving Architectures (ADA) Research Center, a JUMP Center co-sponsored by SRC and DARPA, the NSF under CCF-1704834, and Intel Corporation. 

%
%
%

\clearpage